\documentclass{aa}  

\usepackage{txfonts}
\usepackage{graphicx}
\usepackage{mathrsfs}
\usepackage{siunitx}
\usepackage{hyperref}
\usepackage{xcolor}

\newcommand{\dsp}{\displaystyle}
\newcommand{\beq}{\begin{equation}}
\newcommand{\eeq}{\end{equation}}

\newcommand{\bal}{\begin{align}}
\newcommand{\eal}{\end{align}}
\newcommand{\mltc}[1]{\multicolumn{1}{c}{#1}}

\newcommand\DEM{\ensuremath{\textrm{DEM}}}
\newcommand\EM{\ensuremath{\textrm{EM}}}
\newcommand\LF{\ensuremath{\textrm{LF}}}
\newcommand\HF{\ensuremath{\textrm{HF}}}
\newcommand\phot{\ensuremath{\textrm{P}}}
\newcommand\coro{\ensuremath{\textrm{C}}}

\begin{document}

\title{Measuring relative abundances in the solar corona with optimised linear combinations of spectral lines}
\titlerunning{Coronal abundances with optimised combinations of spectral lines}
\author{Natalia Zambrana Prado\inst{1}, Éric Buchlin\inst{1}}
\institute{Institut d'Astrophysique Spatiale, CNRS/Université Paris-Sud, Université Paris-Saclay, bâtiment 121, Université Paris-Sud, 91405 Orsay cedex, France\\
          \email{natalia.zambranaprado@ias.u-psud.fr}
          }

\date{Received 28 November 2018 \/ Accepted 30 September 2019}

\abstract
{Elemental abundances in some coronal structures differ significantly from photospheric abundances, with a dependence on the first ionization potential (FIP) of the element.
Measuring these FIP-dependent abundance biases is important for coronal and heliospheric physics.}
{We aim to build a method for optimal determination of FIP biases in the corona from spectroscopic observations in a way that is in practice independent from Differential Emission Measure (DEM) inversions.}
{We optimised linear combinations of spectroscopic lines of low-FIP and high-FIP elements so that the ratio of the corresponding radiances yields the relative FIP bias with good accuracy for any DEM in a small set of typical DEMs.}
{These optimised linear combinations of lines allow retrieval of  a test FIP bias map with good accuracy for all DEMs in the map.
The results also compare well with a FIP bias map obtained from observations using a DEM-dependent method.}
{The method provides a convenient, fast, and accurate way of computing relative FIP bias maps. It can be used to optimise the use of existing observations and the design of new observations and instruments.}


\maketitle

\section{Introduction}

In order to understand the interactions between the Sun and the heliosphere and their impact on the celestial bodies living within the latter, we need to study the properties and the origin of the solar wind (SW), which shapes the heliosphere. Accurate plasma diagnostics of the SW and the corona, the uppermost layer of the solar atmosphere, and precise modelling of the solar magnetic field and plasma flows in the interplanetary medium are crucial when trying to determine the source regions of the SW \citep{Peleikis2017}. Indeed, the chemical composition of coronal plasma (the abundances of the different elements) may vary from structure to structure and in time \citep{Feldman03} but it becomes fixed at low heights in the corona. Determination of the composition of the different structures would allow us to pinpoint the source of the SW by comparing and linking remote-sensing abundance measurements to in situ analysis.

Variations in coronal plasma abundances can be found in different types of structures such as active regions \citep{Baker13}, jets, plumes \citep{Guennou2015}, and loops. These variations are linked to the first ionization potential \citep[FIP;][]{Saba1995} of the different elements.  Typically, in magnetically closed structures, the coronal abundances of elements that have a low FIP ($<10$\;eV) are enhanced in comparison to their photospheric abundances. This is not the case for elements with a higher FIP (for these elements the coronal and photospheric abundances are about the same). This anomaly is called the FIP effect \citep{Pottasch64a, Pottasch64b}, and it can be quantified by measuring the ratio of the coronal to photospheric abundance (the abundance bias, also referred to as FIP bias as it is FIP-dependent) of different elements. These anomalies do not only occur in the corona of the  Sun but also in other stellar coronas, and an `inverse FIP effect' has even been detected in some of them \citep{Laming15}. 

Being able to measure the FIP effect by remote sensing and comparing it to the in situ abundance diagnostics of the SW \citep{VonSteiger97} can therefore allow us to determine the origin of the particles that arrive at a spacecraft \citep{Brooks2011}. Having abundance maps produced systematically from all adequate UV observations would then help to obtain a better idea of how the solar wind is formed and how it unfolds in the interplanetary medium.

Different methods exist to determine photospheric abundances with remarkable accuracy even though there have been significant shifts in abundances of certain elements (oxygen in particular) throughout the years \citep{Caffau11, Schmelz12, Grevesse15, Scott2015, Scott15}. Photospheric abundances do not vary with solar surface location or from one particular solar feature to another. On the other hand, coronal abundances, which are derived from UV spectroscopy, are much more difficult to measure accurately, as evidenced by the discrepancies between different measurements that were previously taken as a reference by the solar community \citep{Schmelz12}. Even though the radiance of a UV spectral line emitted by an ion is proportional to its abundance, the latter is difficult to determine. This can be explained because many other parameters come into play, related to plasma conditions or to atomic physics, with high uncertainties for some of them.

First ionization potential biases are usually calculated either from the line ratio of two spectral lines (hereafter 2LR method) or following Differential Emission Measure (DEM) analysis; both these methods can yield different results when used on the same data. In order to accurately obtain a FIP bias with the 2LR technique, both spectral lines have to be formed at very close temperatures, while using the DEM allows more flexibility in the choice of lines. However, the DEM is difficult to estimate accurately \citep{Craig76, Judge1997, Landi2011, Testa2012, Guennou2012b}, especially when trying to design an automated method.

In this paper we present a new method, developed with the aim to provide optimal determination of the abundance biases in the corona from a spectroscopic observation, even when the DEM cannot be precisely determined. The method is based on the DEM formalism and relies on linear combinations of spectral lines to get rid of the dependence on DEM inversion for FIP bias determination. We can think of several uses of such a method:
\begin{itemize}
    \item obtaining FIP bias maps from an existing observation that had not been specifically designed for this purpose;
    \item designing an observation to obtain the best FIP bias map possible with a given instrument;
    \item releasing the constraints on the list of spectroscopic lines required to build FIP bias maps, allowing the design of observations that perform more diagnostics simultaneously;
    \item ultimately, helping to design the next solar or stellar UV spectrometers with FIP bias measurement capability.
\end{itemize}

In Sect.~\ref{sec:spectroscopy} we present the theoretical background for the method, which we present in Sect.~\ref{sec:our_method}.
We test its accuracy and DEM independence and compare its results to those obtained by the line ratio technique using synthetic spectra in Sect.~\ref{sec:tests}.
We then apply our method to existing active region spectra obtained with the Hinode/EIS spectrometer and previously studied by \cite{Baker13} in Sect.~\ref{sec:baker_lines}.
We finish by discussing our results and the interest of the method in Sect.~\ref{sec:discussion}, where we also present our conclusions.

\section{Theoretical background of FIP bias measurements in the corona}
\label{sec:spectroscopy}

\subsection{Contribution functions and differential emission measure}
\label{sec:contrib}

In the tenuous and hot corona \citep[in the so-called coronal approximation, ][]{Mason1994,Landini1990}, the radiance of an optically thin spectral line at wavelength $\lambda_{ij}$ corresponding to the transition $j \to i$ of the $X^{+m}$ ions can be written as
\begin{equation} \label{eq:intens}
    I_{ij} = \frac{1}{4\pi} \int N(X^{+m}_j) \, A_{ij} \, h \nu_{ij} \; dz,
\end{equation}
where $\nu_{ij}$ is the frequency corresponding to the transition, $A_{ij}$ is the Einstein coefficient for spontaneous emission, $N(X_{j}^{+m})$ is the density of $X^{+m}$ ions in level $j$, and integration is over the line of sight.

In order to evaluate the quantities involved in Eq.~\eqref{eq:intens} from the plasma parameters and atomic physics, this equation is frequently rewritten as
\begin{equation}\label{eq:int+cofnt}
    I_{ij} = \int A_X^\coro \, C_{ij}(T, N_e) \, N_e^2 \; dz,
\end{equation}
where $\dsp{A_X^\coro \equiv \frac{N(X)}{N(H)}}$ is the elemental abundance in the corona relative to hydrogen, $N_e$ is the electron density, and $C_{ij}(T, N_e)$ is the contribution function for the spectral line. This contribution function can be rewritten, in the simple case of a two-level ion, as
\begin{equation}\label{eq:cofnt}
    C_{ij}(T, N_e) \equiv \frac{h \nu_{ij}}{4\pi} \frac{N(X_{j}^{+m}) \, A_{ij}}{N(X^{+m}) \, N_e} \frac{N(X^{+m})}{N(X)} \frac{N(H)}{N_e}.
\end{equation}
In this equation, we recognise the relative level $j$ population of ion $X^{+m}$, the relative population of ionization stage $+m$ of element $X$, and the hydrogen abundance relative to free electrons $N(H)/N_e$. The latter is usually taken as $0.83$ in the corona as hydrogen and helium are almost completely ionized at $T > 10^5\;\mathrm{K}$.

The contribution function contains all atomic physics parameters that play a role in line formation, and it is different for each spectral line. In the case of many-level ions, Eq.~\eqref{eq:int+cofnt} is still valid, but $C_{ij}$ has to be computed through more complex atomic physics models.
In contrast, the distribution of $N_e^2$ as a function of temperature along the line of sight is the same for all lines.
However, as the spatial distribution of the plasma parameters $T$ and $N_e$ along the line of sight is lost in integration, while the usual shape of contribution functions (with a strong dependence on temperature and a weaker dependence on density for most spectral lines) tends to `select' some temperature range for a given spectral line, it is useful to substitute $T$ for $z$ in the integral, by writing
\begin{equation}\label{eq:int+cofnt+dem}
    I_{ij} = \int A_X^\coro \, C_{ij}(T, N_e) \, \DEM(T) \; dT,
\end{equation}
where the DEM can be defined by
\begin{equation}
    \DEM(T) = N_e^2 \frac{dz}{dT}
\end{equation}
in the simple case where temperature is a strictly monotonous function of the position along the line of sight.

Different methods exist to determine the DEM from the observed radiances in several spectroscopic lines using integral inversion methods. However, this is not an easy task as this method has many limitations \citep{Craig76, Laming15, Landi2011}. For most DEM determination techniques, a previous measurement of the density is needed. The insufficiency of the available data as well as the intrinsic nature of DEM inversion make it a difficult, ill-constrained problem that is very poorly conditioned in the density dimension \citep{Judge1997, Testa2012}. Through the application of some of these inversion methods to synthetic data, one finds that the general shape of the DEM is not always well retrieved and the finer details are not always well resolved \citep{Testa2012}. As in any such problem, different $\DEM(T)$ functions can equally reproduce the observed radiances. Furthermore, when dealing with synthetic observations of multithermal plasma, DEM inversion fails to find a good match with the `true' DEM \citep{Testa2012} and isothermal DEM inversion solutions for a multithermal plasma are biased to specific temperature intervals, for a given set of spectroscopic lines \citep{Guennou2012b}.

We now take a quick look at the current FIP bias-determination methods based on this formalism. 

\subsection{First ionization potential bias determination}\label{sec:spec+FIP}

Let us consider two spectroscopic lines emitted by ions of two different elements: $X_\LF$ that has a low FIP (LF, $<10\,\mathrm{eV}$) and $X_\HF$ that has a high FIP (HF). The radiance of the considered
spectral line of the low-FIP and high-FIP elements is denoted $I_\LF$ and $I_\HF$ , respectively.

Assuming that abundances are uniform along the relevant part of the line of sight, in the corona, we can write Eq.~\eqref{eq:int+cofnt+dem} for both lines as
\begin{align}
    I_\LF &= A_{X_\LF}^\coro \langle C_\LF, \DEM\rangle \label{eq:int=sum+FIP_LF}\\
    I_\HF &= A_{X_\HF}^\coro \langle C_\HF, \DEM\rangle,  \label{eq:int=sum+FIP_HF}
\end{align}
where $A^\coro_X$ are the coronal abundances for each element, $C_\LF$ and $C_\HF$ are the contribution functions for the lines of the low-FIP and high-FIP elements, respectively, and $\langle a, b \rangle \equiv \int a(T) \, b(T) \; dT$ is a scalar product.

Introducing the photospheric abundance $A_X^\phot$ and the FIP bias $f_X \equiv A_X^\coro / A_X^\phot$ for element $X$, the ratio of line radiances becomes

\begin{equation}\label{eq:lineratio}
\frac{I_\LF}{I_\HF} = \frac{A_{X_\LF}^\phot}{A_{X_\HF}^\phot} \frac{f_{X_\LF}}{f_{X_\HF}}
\frac{\langle C_\LF, \DEM\rangle}{\langle C_\HF, \DEM\rangle}.
\end{equation}

When the DEM can be inferred from observations and the contribution functions computed from atomic calculations, the relative FIP bias between high-FIP and low-FIP elements can then be derived from Eq.~\eqref{eq:lineratio} using the observed radiances and assuming the photospheric abundances:
\begin{equation}\label{eq:fipdem}
\frac{f_{X_\LF}}{f_{X_\HF}} = \frac{I_\LF}{I_\HF}
\left(
    \frac{A_{X_\LF}^\phot}{A_{X_\HF}^\phot} 
    \frac{\langle C_\LF, \DEM\rangle}{\langle C_\HF, \DEM\rangle}
\right)^{-1}.
\end{equation}
This ratio is simply the low-FIP element abundance bias $f_{X_\LF}$ if we consider that $f_{X_\HF}=1$.

In practice, the required DEM inversion itself is sensitive to FIP bias, especially because DEM inversion often involves iron lines, a low-FIP element.
This sensitivity can however be used as a way to determine the FIP bias, as in e.g. \citet{Baker13,Guennou2015}.
A short step-by-step description of their process is presented in Appendix~\ref{app:dem_inversion}.

Another method for FIP bias determination is the 2LR method, which does not involve DEM inversion.
When two spectral lines, one from a low-FIP ion and another from a high-FIP ion, can be chosen so that their contribution functions are very close (at some factor which can then be approximated by $\max(C_\LF) / \max(C_\HF)$), the ratio of the scalar products in Eq.~\eqref{eq:lineratio} becomes almost independent from the DEM, and the relative FIP bias becomes
\begin{equation}
    \frac{f_\LF}{f_\HF} \approx \frac{I_\LF}{I_\HF}
    \left(
        \frac{A_{X_\LF}^\phot \max(C_\LF)}{A_{X_\HF}^\phot  \max(C_\HF)}\right)^{-1}.
        \label{eq:ratioscal}
\end{equation}
This is simply the ratio of the radiances multiplied by some constant factor.

Of course, finding such adequate line pairs of low-FIP and high-FIP elements with similar contribution functions is difficult and not always possible given the observational constraints.
Furthermore, no two contribution functions are exactly the same, so there is some hidden dependence on the DEM, and this method is then less accurate than using Eq.~\eqref{eq:fipdem} after inversion of the DEM.

However, as mentioned above, DEM inversion is a difficult problem, and therefore a FIP bias determination that would not rely on DEM inversion, like the 2LR method, but that would also be more accurate than the two-line ratio method would be very convenient.
This is the main motivation for the development of our method.

\section{A new way of measuring the FIP effect: the linear combination ratio method} \label{sec:our_method}

\subsection{Light bulb}

For some elements, as mentioned above, we can make relative abundance diagnostics without knowing the DEM, by using radiance ratios of the 
spectral lines of a low-FIP element and a high FIP element, provided they both have very similar contribution functions.
Such lines are not always observable however, or their contribution functions are not close enough.
Our idea is therefore to generalise this technique by using linear combinations of lines so that the corresponding contribution functions for low-FIP and high-FIP elements are a better match.

We start by defining two radiance-like quantities that would be the analogues of the radiances of Eqs.~\eqref{eq:int=sum+FIP_LF}--\eqref{eq:int=sum+FIP_HF}, as linear combinations of radiances from individual lines of low-FIP and high-FIP elements:
\begin{align}
    \label{eq:pseudo_intensity}
    \mathscr{I}_\LF &\equiv \sum_{i \in (\LF)} \alpha_i \; \frac{I_i}{A^\phot_i}  ,\\
    \mathscr{I}_\HF &\equiv \sum_{i \in (\HF)} \beta_i \; \frac{I_i}{A^\phot_i}.
\end{align}
Please note that the normalization by photospheric abundances here is only a matter of convention.

Using Eq.~\eqref{eq:int+cofnt+dem}, these quantities become
\begin{align}
    \mathscr{I}_\LF &= \sum_{i \in (\LF)} \alpha_i \; f_i \; \langle C_i, \; \DEM \rangle,\\
    \mathscr{I}_\HF &= \sum_{i \in (\HF)} \beta_i \; f_i \; \langle C_i, \; \DEM \rangle.
\end{align}

If the FIP biases of all used low-FIP elements are the same (and equal to $f_\LF$), and the FIP biases of all used high-FIP elements are the same (and equal to $f_\HF$), the ratio of the radiance-like quantities is
\begin{align}
    \frac{\mathscr{I}_\LF}{\mathscr{I}_\HF} &= \frac{f_\LF \sum_{i \in (\LF)} \alpha_i \; \langle C_i, \DEM \rangle}{f_\HF \sum_{i \in (\HF)} \beta_i\;  \langle C_i, \; \DEM \rangle} \\ 
    &= \frac{f_\LF \; \langle \mathscr{C}_\LF, \DEM \rangle}{f_\HF \;  \langle \mathscr{C}_\HF, \DEM \rangle} \label{eq:lincombradratio},
\end{align}
where the low-FIP and high-FIP contribution functions have been defined by
\begin{align}
    \mathscr{C}_\LF(T) &\equiv \sum_{i \in (\LF)} \alpha_i \; C_i(T) \label{eq:lc1},\\
    \mathscr{C}_\HF(T) &\equiv \sum_{i \in (\HF)} \beta_i \; C_i(T), \label{eq:lc2}
\end{align}
then the relative FIP bias is
\begin{equation} \label{eq:biaslc}
    \frac{f_\LF}{f_\HF} = \frac{\mathscr{I}_\LF}{\mathscr{I}_\HF}
        \left(
            \frac{\langle \mathscr{C}_\LF, \DEM \rangle}{\langle \mathscr{C}_\HF, \; \DEM \rangle}
        \right)^{-1} .
\end{equation}
This is analogous to Eq.~\eqref{eq:ratioscal} but for the linear combinations of radiances and of contribution functions.

\subsection{Finding the optimal linear combinations}

So that the relative FIP bias can be retrieved from observations without determining the DEM, our first idea was to optimise the linear combination coefficients so that the following cost function is minimised:
\begin{equation}
    \phi \, (\alpha, \beta)
    = \left\lVert\mathscr{C}_\LF - \mathscr{C}_\HF \right\rVert
    = \left\lVert \sum_{i \in (\LF)} \alpha_i C_i - \sum_{i \in (\HF)} \beta_i C_i \right\rVert
    \label{eq:costcontrib}
,\end{equation}
where the distance is defined from the scalar product: $||a||^2 \equiv \langle a, a \rangle$.
If there is no difference between $\mathscr{C}_\LF$ and $\mathscr{C}_\HF$, the relative FIP bias from Eq.~\eqref{eq:biaslc} is indeed simply $\mathscr{I}_\LF / \mathscr{I}_\HF$.
However, if differences remain, especially in the wings of the linear combinations of contribution functions, the result remains sensitive to the DEM.

We therefore decided to look at the problem from a different angle but with a similar approach. Instead of building the cost function $\phi$ from the distance between the contribution functions as in Eq.~\eqref{eq:costcontrib}, we came back to Eq.~\eqref{eq:lincombradratio} and built a new cost function in such a way that after optimisation the ratio
$\langle \mathscr{C}_\LF, \DEM \rangle / \langle \mathscr{C}_\HF, \DEM \rangle$
would become as close to 1 as possible for {any} DEM.
As we do not want to compute the DEM in each pixel, this means that we have to choose a set $(\DEM_j)_j$ of `reference' DEMs that would be representative of the DEMs in the map, and we then define the cost function as
\begin{align}
    \phi \, (\alpha, \beta) &=
    \sqrt{\sum_{j\in(\DEM_j)_j} \left | \frac{\sum_{i \in (\LF)} \alpha_i \; \langle C_i, \DEM_j \rangle}{\sum_{i \in (\HF)} \beta_i\;  \langle C_i, \; \DEM_j \rangle}-1 \right |^2}.
    \label{eq:costrad}
\end{align}
This is simply the L2 distance between vector
\begin{equation}
    \label{eq:costvec}
    (\psi_j)_j \equiv (\langle \mathscr{C}_\LF, \DEM_j \rangle / \langle \mathscr{C}_\HF, \DEM_j \rangle)_j
\end{equation}
and vector $(1)_j$.

Through the minimisation of $\phi$ we obtain the coefficients $\alpha_i$ and $\beta_i$.
Provided that the set of DEMs used to define the cost function from Eq.~\eqref{eq:costrad} is adequate, we then have $\psi_\DEM = \langle \mathscr{C}_\LF, \DEM \rangle / \langle \mathscr{C}_\HF, \DEM \rangle \approx 1$ in each pixel and following Eq.~\eqref{eq:biaslc} the relative FIP bias can be simply retrieved as
\begin{equation} \label{eq:biaslcopt}
    \frac{f_\LF}{f_\HF} = \frac{\mathscr{I}_\LF}{\mathscr{I}_\HF} \, \psi_\DEM^{-1}  \approx \frac{\mathscr{I}_\LF}{\mathscr{I}_\HF}.
\end{equation}
This allows us to build relative FIP bias maps from spectroscopic observations without having to determine the DEM in each pixel.
We call this method Linear Combination Ratio (LCR) method.

We note that the 2LR technique, as defined by Eq.~\eqref{eq:ratioscal}, is a special case of a linear combination and can be expressed with the same formalism as the LCR method:
a single line from a low-FIP element is chosen along with a single line from a high-FIP element, and the linear combination coefficients are
\begin{align}\label{eq:2LR_coeffs}
    \alpha = \frac{1}{\max(C_\LF)} \qquad \textrm{and} \qquad
    \beta = \frac{1}{\max(C_\HF)}
\end{align}
for the single \LF\ and \HF\ lines, instead of being the result of an optimisation.

\subsection{Implementing the LCR method}\label{subsec:implementation}

We have developed a Python module\footnote{\url{https://git.ias.u-psud.fr/nzambran/fiplcr}} to compute the optimal linear combinations of spectral lines and to use them to compute relative FIP bias maps from observations.

We describe the different steps to apply the method to observations of UV spectra in the following.

\subsubsection{Selection of the spectral lines}\label{subsubsec:line_selection}

We first need to choose the spectral lines that we want to use.
This has to be done by hand, and depends on the lines available for a given observation, instrument, or wavelength range.
The following criteria should be taken into consideration:
\begin{itemize}
    \item The lines have to verify the coronal approximation (see Sect.~\ref{sec:contrib}).
    \item They have to form at coronal temperatures; if they form at lower temperatures opacity effects would have to be taken into account.
    \item The observed signal-to-noise ratio of each line radiance has to be sufficient.
    The noise in the observed radiance of a weak line propagates indeed to the corresponding linear combination of radiances, especially when it is amplified by a large coefficient in the linear combination.
    \item Blended lines have to be avoided or de-blended so that the true spectral line radiance is used.
\item The atomic physics for the spectral lines has to be well known. Examples of some of the problems that one might encounter are an underestimation of the observed flux in the 50–130\,\AA\ wavelength range \citep{Testa2012b} or anomalous behaviour for ions of the Li and Na isoelectronic sequences for which the atomic physics models tend to underestimate line radiance (\citealt{DelZanna2001}; Sect.~7.4 of \citealt{delzannag18a}). The quality of the atomic data will depend on the database chosen. For the purpose of this paper, we use CHIANTI (version 9.0, previous versions described in \citealt{Dere97}, \citealt{DelZanna2015}) through the \texttt{ChiantiPy} Python package (version 0.8.5), but in principle another database can be used.
    \item The maximums of the contribution functions of all lines should be at similar temperatures so that we do not mix abundances at various heights.
\end{itemize}

\subsubsection{Computation of the contribution functions}\label{sec:computation_cofnt}

We use the CHIANTI atomic physics database to compute the contribution functions. We also use it to retrieve information about each spectral line, such as typical photospheric abundance, FIP of the element, and upper and lower levels of the transition.
Furthermore, for the cases when density maps can be obtained, for example when using a radiance ratio between a pair of lines with different density sensitivities, we compute these contribution functions on a grid of temperatures and densities.

\subsubsection{Determination of the optimal linear combinations}
\label{sec:optimlc}

Lines are first  separated into two subsets, $(\LF)$ for lines from low-FIP elements and $(\HF)$ for lines from high-FIP elements. We then minimise the cost function $\phi$ defined by Eq.~\eqref{eq:costrad} using a very simple set of DEMs, constituted by the typical DEMs provided by CHIANTI for an active region (AR), a coronal hole (CH), and the quiet Sun (QS).
For this minimisation, we use the \citet{NelderMead1965} optimisation implemented in the \texttt{SciPy} library \citep{scipy}.
As a first guess for the coefficients for each linear combination, we use the median of the maximums of the contribution functions divided by each of these maximums.
The optimisation then yields a set of optimal coefficients, $(\alpha_i)_{i\in \LF}$ and $(\beta_i)_{i\in \HF}$. These coefficients can be optimised for the density grid mentioned in Sect.~\ref{sec:computation_cofnt} if required; we would then obtain optimised linear combination coefficients that are a function of density.

\subsubsection{Determining the relative FIP bias}

Once we have the coefficients of the linear combinations, we can compute the linear combinations of radiances $\mathscr{I}_\LF$ and $\mathscr{I}_\HF$ in each pixel, and then immediately obtain the relative FIP bias $f_\LF / f_\HF$ from Eq.~\eqref{eq:biaslcopt}.
If a density map can be obtained for the observation, we can compute the FIP bias in each pixel using the linear combination coefficients best suited for that particular pixel depending on its density (the density dependence of the ratio of the radiances is further discussed in Appendix~\ref{app:density_independance}).

\section{Testing the method with synthetic radiances} \label{sec:tests}

\newlength\highfigwidth
\makeatletter
\ifaa@referee
    \setlength{\highfigwidth}{0.7\linewidth}
\else
    \setlength{\highfigwidth}{\linewidth}
\fi
\makeatother

\begin{figure}[ht]
    \centering
    \includegraphics[width=.8\highfigwidth]{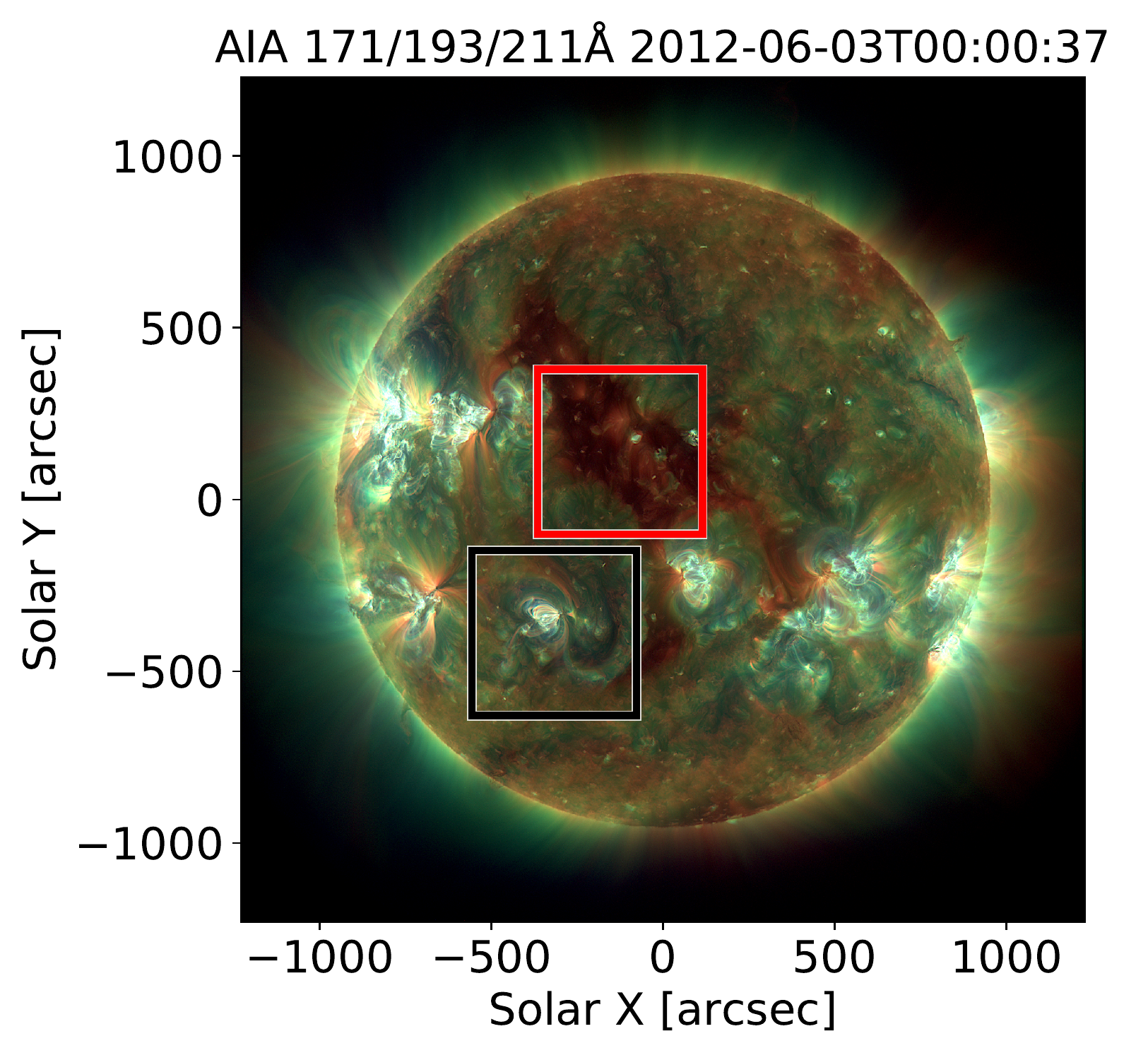}%
    \vspace{-3ex}\\
    \includegraphics[width=.8\highfigwidth]{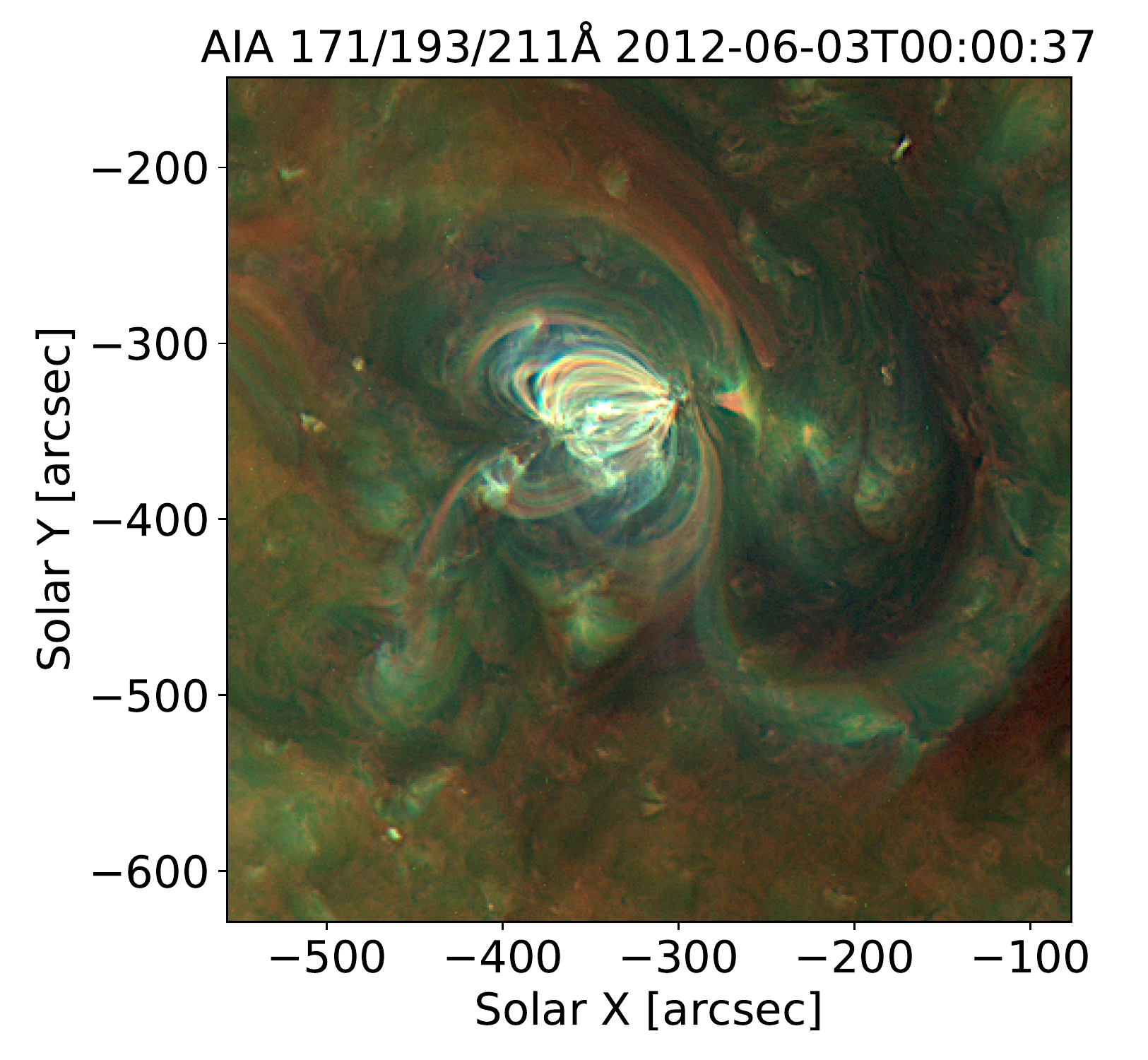}%
    \vspace{-3ex}\\
    \includegraphics[width=.8\highfigwidth]{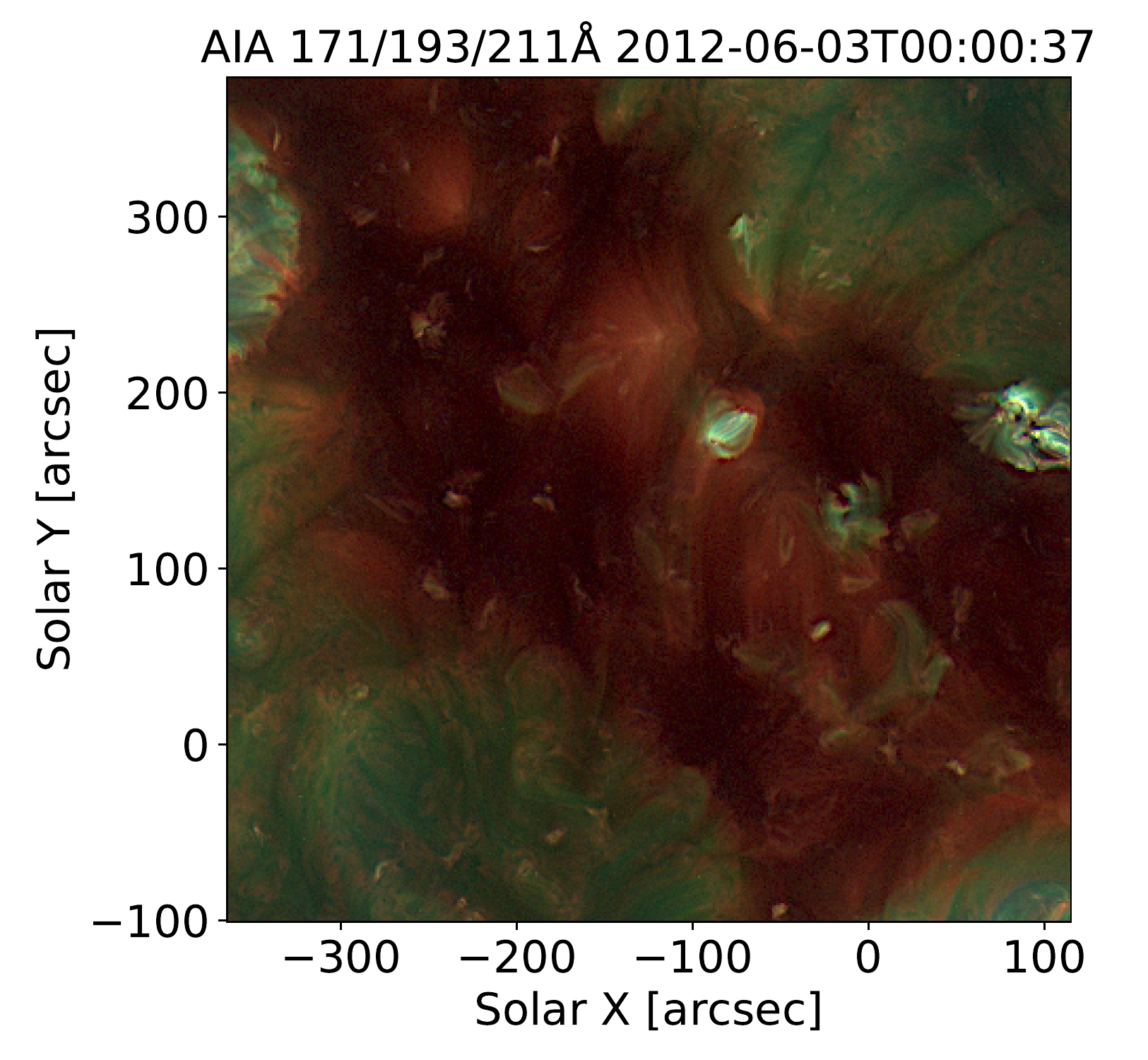}
    \caption{Top panel: Composite map of the solar corona on June 3,\textsuperscript{} 2012, in the 171\,\AA\ (red), 193\,\AA\ (green), and 211\,\AA\ (blue) channels of the AIA instrument aboard SDO. The black and red squares correspond to the regions of interest used for testing the method, and are centred on an AR and a CH, respectively. Middle panel: Zoom on the AR region of interest (black square in top panel). Bottom panel: Zoom on the CH region of interest (red square in top panel).}
    \label{fig:aia}
\end{figure}

\begin{figure}
    \includegraphics[width=\linewidth]{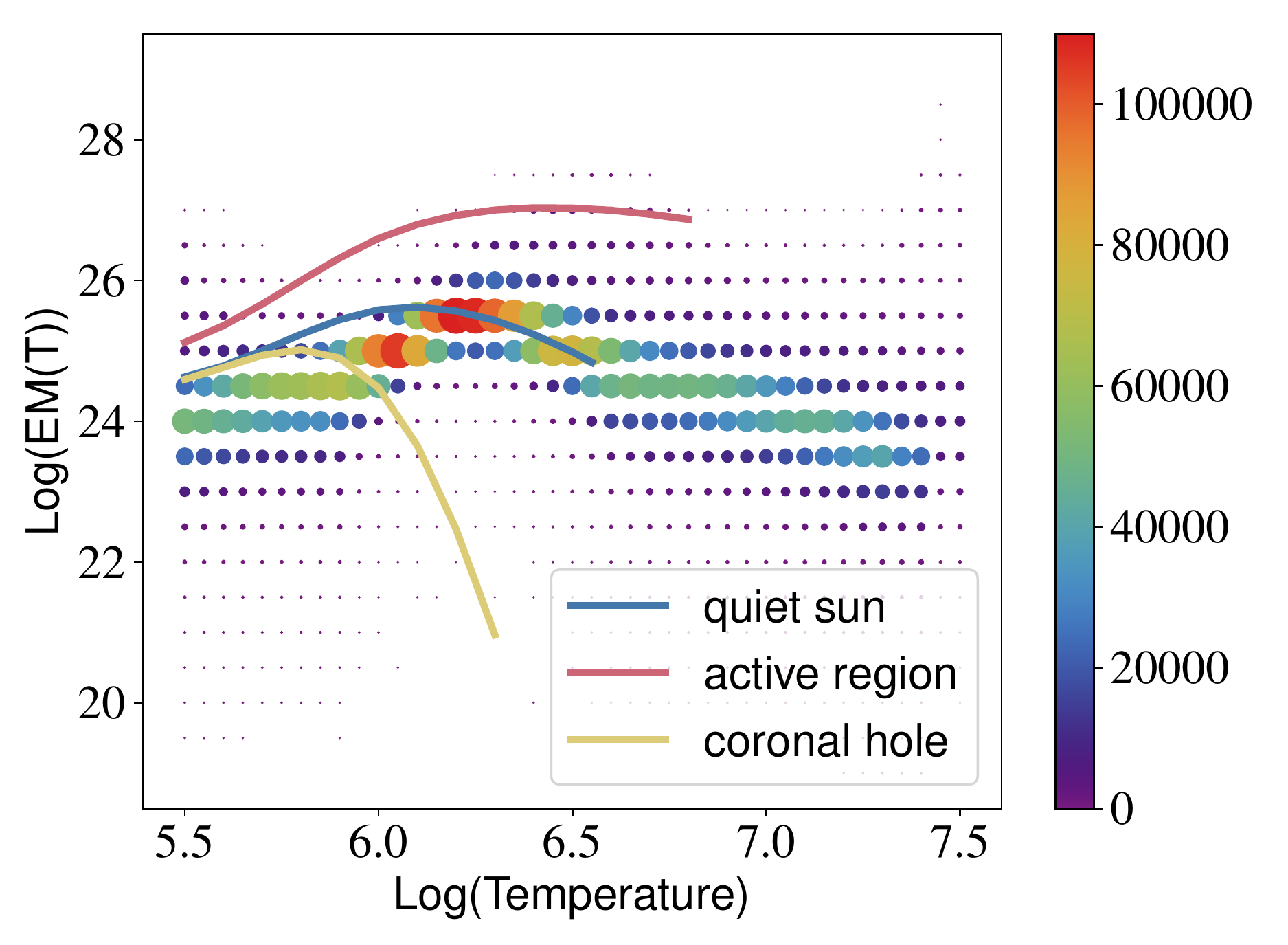}
    \caption{Histogram of the EM($T$) values in the first region of interest (black square) of Fig.~\ref{fig:aia}. The colour scale and the size of the points correspond to the number of pixels containing an EM of a given value at a given temperature. We have also traced in full lines the typical EMs from CHIANTI that we use to optimise the cost function (Eq.~\ref{eq:costrad}). The red line corresponds to an active region, the blue line to the quiet sun and the yellow line to a coronal hole.}
    \label{fig:em_histo_ar}
\end{figure}

\begin{figure}
    \includegraphics[width=\linewidth]{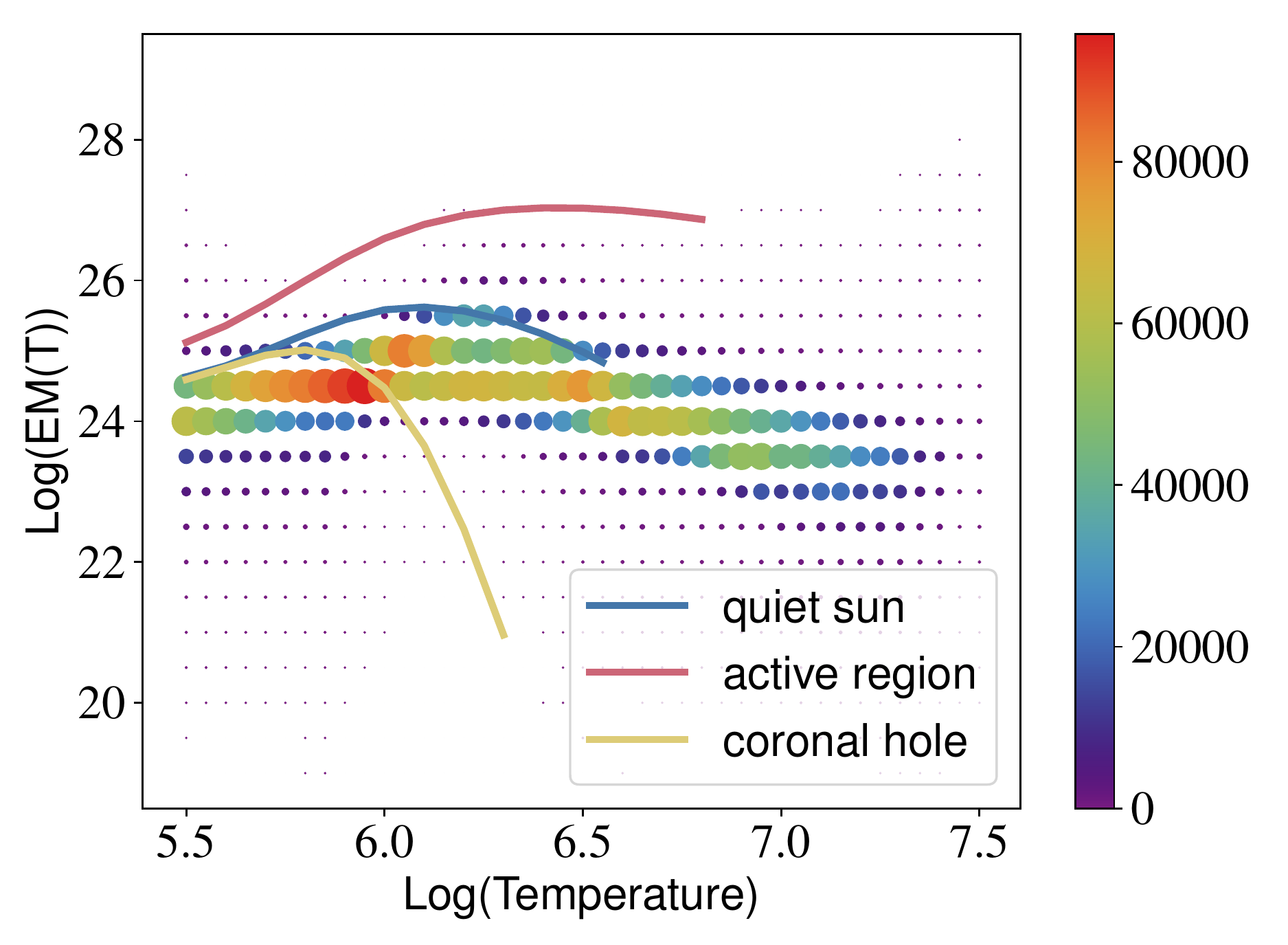}
    \caption{Same as Fig.~\ref{fig:em_histo_ar} but for the second region of interest (red square) of Fig.~\ref{fig:aia}.}
    \label{fig:em_histo_ch}
\end{figure}

In this section we test the LCR method by applying it to maps of synthetic radiances, so that we control all parameters precisely (in particular the abundances).
We also test the 2LR method with the same criteria for comparison.

The test case consists in a uniform abundance map for any given element, combined with a data cube of DEMs, as detailed below.
Using both these inputs and atomic physics, we can build `synthetic' radiances, meaning that they are computed rather than observed.
The test is considered successful for a given FIP bias determination method if the output relative FIP bias map is consistent with the input elemental abundance maps, both in uniformity and in value.
The test has four main steps, detailed below:
\begin{enumerate}
    \item We derive a DEM cube from the AIA observation. This is for the sole purpose of producing synthetic radiances, for which we have control over all parameters, while the DEMs are representative of different real solar regions.
    \item Using CHIANTI for the contribution functions and the  derived DEMs, we calculate the synthetic radiances. We assume different uniform abundances for different elements.
    \item We determine the optimal linear combination coefficients for the LCR methods, and the coefficients for the 2LR method.
    \item We use these coefficients to retrieve the FIP bias in each pixel assuming Eq.~\eqref{eq:biaslcopt} is verified. If this is the case, the retrieved FIP bias map should be uniform.
\end{enumerate}

\subsection{Synthetic radiance maps}

We start by computing Emission Measure\footnote{The EM is the DEM($T$) integrated over temperature bins, of width assumed to be $\Delta\log T = 0.05$ in this section. Then $\langle \mathscr{C}, \DEM\rangle \approx \sum_i \mathscr{C}(T_i)\, \EM(T_i)$, where the sum is over the temperature bin centres.} (EM) maps using the \citet{Cheung15} code (version 1.001), which is available in the SDO/AIA package in SolarSoft.
This particular EM inversion method finds a sparse solution, that is, it uses a minimum number of basis functions to produce an $\EM(T)$ compatible with the observations.

We chose an observation of the Atmospheric Imaging Assembly \citep[AIA;][]{Lemen12} instrument aboard the Solar Dynamics Observatory \citep[SDO;][]{Pesnell12}, on June 3, 2012, close to the maximum of solar activity, so that the coronal emission is inhomogeneous.
The Sun in this particular day presented various ARs and a large CH at the centre.
As shown in Fig.~\ref{fig:aia}, we select two separate regions of interest, one being centred on an AR, and the other centred on a CH.
We aligned the images from the different detectors and divided each of them by their corresponding exposure time before doing the EM computations. 
The histograms of the EMs we obtain are shown in Figs.~\ref{fig:em_histo_ar} and~\ref{fig:em_histo_ch}.

\begin{table}
\caption{Spectral lines used to perform the calculations. All lines were used for the LCR method in Sect.~\ref{sec:tests} and~\ref{sec:baker_lines}. The lines in bold correspond to those used for the 2LR method in the tests in Sect.~\ref{sec:tests}.
For the 2LR method, the coefficients are defined by Eq.~\eqref{eq:2LR_coeffs} which explains the factor $10^{25}$ by which both values are multiplied. For the LCR method they result from the optimisation. These coefficients were calculated for a density of log(n) = 8.3.}
\label{tab:spec_lines}
\begin{center}
\begin{tabular}{lS[table-format=3.3]SS[table-format=1.4]S}
\hline\hline
    Ion                     & \mltc{Wavelength}     & \mltc{$\log T_\textrm{max}$}   & \mltc{LCR coeff}  & \mltc{2LR coeff} \\
    & \mltc{(\AA)} & \mltc{(K)} & & \mltc{($10^{20}$)} \\ \hline
    \ion{Fe}{xii}           & 195.119               & 6.2   & 0.0845      &  \\
    \ion{Fe}{xiii}          & 201.126               & 6.2   & -0.0738     &  \\
    \ion{Fe}{xiii}          & 202.044               & 6.2   & 0.0294      &  \\
    {\bfseries \ion{Si}{x}} & {\bfseries 258.374}   & 6.1   & 1.36        &  4.26\\
\ion{Si}{x}             & 261.056               & 6.1   & 1.46        &  \\
    {\bfseries \ion{S}{x}}  & {\bfseries 264.231}   & 6.2   & 2.16        &  3.34  \\
    \ion{Fe}{xiv}           & 264.789               & 6.3   & 0.503       &  \\
    \ion{Fe}{xiv}           & 274.204               & 6.3   & 0.0404       &  \\
\hline
 \end{tabular}
 \end{center}
\end{table}

\begin{table}
\caption{First ionization potential of the elements used for the tests, their coronal and photospheric abundances taken from \cite{Schmelz12} and \cite{Grevesse2007}, and the corresponding abundance bias relative to sulfur.}
\label{tab:FIP_bias_ratios}
\begin{center}
\begin{tabular}{lSccc}
\hline\hline
Element & \mltc{FIP (eV)}   & $A^\coro_\mathrm{X}$      & $A^\phot_\mathrm{X}$  & $f_X/f_\mathrm{S}$  \\ \hline
Fe      & 7.90              & $7.08\times 10^{-5}$     & $2.82\times 10^{-5}$ & 2.05                \\
Si      & 8.15              & $7.24\times 10^{-5}$     & $3.24\times 10^{-5}$ & 1.82                \\
S       & 10.36             & $1.69\times 10^{-5}$     & $1.38\times 10^{-5}$ & 1.00                \\ \hline
\end{tabular}
\end{center}
\end{table}

We  selected lines available in the observations used by \citet{Baker13}, as in Sect.~\ref{sec:baker_lines} we use the same observational data as these authors to compare the LCR method results to their results. Eight lines were chosen following the criteria of Sect.~\ref{subsubsec:line_selection} and are listed in Table~\ref{tab:spec_lines}. The temperature range of their maximums of formation goes from 1 MK to 2 MK.
They include five iron lines, two silicon lines, and one sulfur line; iron and silicon are low-FIP elements, and, like \citet{Baker13}, we consider  sulfur to be high-FIP.
We further select \ion{Si}{x} 258.374\,\AA\ as low-FIP line for the 2LR method.

We then create the required synthetic radiance maps using Eq.~\eqref{eq:int+cofnt+dem}, assuming the relative abundance ratios presented in Table \ref{tab:FIP_bias_ratios}, which provide the `ground truth' for the FIP biases we obtain using both methods.
The abundances we assume here are uniform\footnote{Uniformity allows an easy comparison between the obtained FIP bias maps and the ground truth; however,
any map could be assumed for the test, as the test (from synthetic radiances to FIP bias maps) gives a result that is proportional to the initial FIP bias in each pixel, as long as all the \LF\ or \HF\ element abundance vectors are collinear.} over the regions of interest, and we take their values from \cite{Schmelz12} for the corona and  \cite{Grevesse2007} for the photosphere; these values and resulting relative FIP biases are presented in Table~\ref{tab:FIP_bias_ratios}.

\subsection{Optimisation of the linear combinations of lines}

\begin{figure}[h]
    \includegraphics[width=\linewidth]{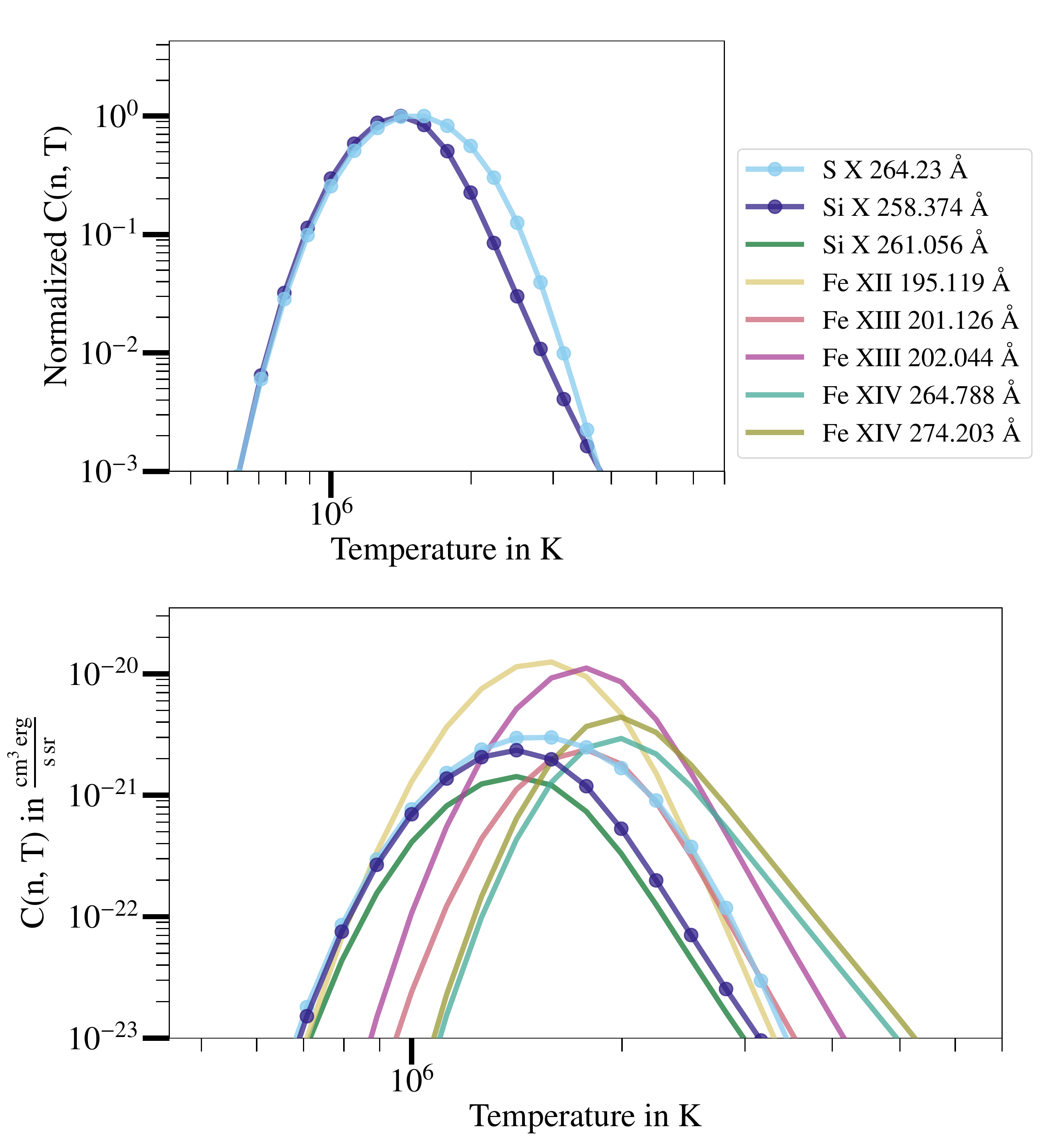}
    \caption{Top: Normalized contribution functions of the lines used for the 2LR method. Bottom: Contribution functions of the lines used for the LCR method. All the contribution functions were calculated assuming a constant density of $10^{8.3} \, \mathrm{cm}^{-3}$.}
    \label{fig:cofnts}
\end{figure}

We present in Fig.~\ref{fig:cofnts} the contribution functions for the spectral lines listed in Table~\ref{tab:spec_lines}. All contribution functions were computed assuming a density of $10^{8.3} \, \mathrm{cm}^{-3}$. In the top panel of Fig.~\ref{fig:cofnts}, we show the contribution functions of both lines used for the 2LR method, normalized by their maximum. As we can see, they are similar in shape from low temperatures until the maximum of both functions, but for higher temperatures they start to differ from one another significantly. In the bottom panel of Fig. 4, we present the contribution functions of all the lines we use to test the LCR method. They all have different shapes and values. Not all of them start at low temperatures, as ions with a high degree of ionization are formed only at higher temperatures.

After choosing the lines and computing their contribution functions, we determine the optimal linear combination of these lines for the LCR method (Sect.~\ref{sec:optimlc}). The reference EMs that we use for the optimisation are plotted on Figs.~\ref{fig:em_histo_ar} and~\ref{fig:em_histo_ch}. These are available in the CHIANTI database and correspond to typical EMs for a coronal hole, an active region, and the quiet sun.
The resulting coefficients are included in Table~\ref{tab:spec_lines}.
For the 2LR method, we use the inverse of the maximum of the contribution functions of the \ion{Si}{x} and of the \ion{S}{x} lines as values for the (single) $\alpha$ and $\beta$ coefficients, respectively, therefore allowing the use of the same formalism as for the LCR method.

\begin{table}
\caption{Value of $\psi_j$ (Eq.~\ref{eq:costvec}) for both methods for each reference DEM, as well as the resulting cost function $\phi$ (Eq.~\ref{eq:costrad}).}
\label{tab:vector_obj_func}
\begin{tabular}{ccccc}
\hline\hline
    Method  & $\psi_\mathrm{QS}$    & $\psi_\mathrm{AR}$    & $\psi_\mathrm{CH}$    & $\phi$ \\ \hline
    2LR     & 0.882              & 0.763              & 1.13              & 0.295  \\
    LCR     & 0.99998              & 1.00000              & 1.00001              & $2.8 \times 10^{-5}$ \\
\hline
 \end{tabular}
\end{table}

In particular, we can compute the cost function defined in Eq.~\eqref{eq:costrad} for both the LCR method (following optimisation) and the 2LR method, as shown in Table~\ref{tab:vector_obj_func}.
In this table we also give the components of vector $\psi$ defined in Eq.~\eqref{eq:costvec}, which ideally would all have to be equal to 1 so that the cost function $\phi$ would be zero.
The values in this table already show that the optimisation made in the LCR method yields much better values for the cost function, as well as for each of the $\psi$ components, compared to the same quantities for the line coefficients chosen for the 2LR method.
This means that Eq.~\eqref{eq:biaslc} would give very good estimates of the relative FIP bias for any of the three reference DEMs that we use.
It is a first indication that the LCR method could work well.

\subsection{First ionization potential bias maps obtained from the synthetic radiances}
\label{subsec:synthetic}

Applying Eq.~\eqref{eq:biaslc} to the synthetic radiance maps, we now obtain maps of the relative FIP bias for both LCR and 2LR methods.

\begin{figure}
    \includegraphics[width=\linewidth]{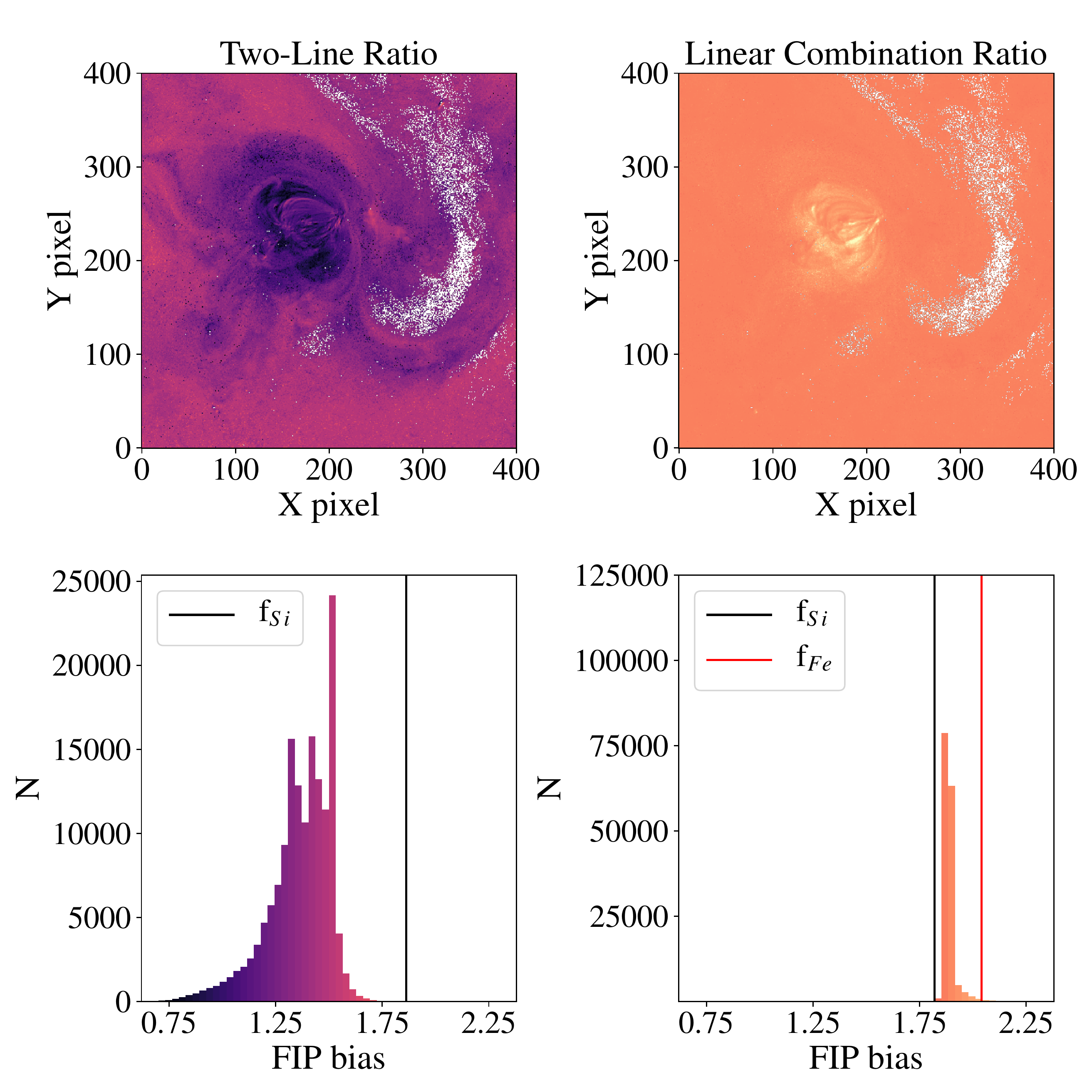}
    \caption{Results of FIP bias determination using the 2LR (left) and LCR (right) methods on the synthetic radiances in the first region of interest (black square) of Fig.~\ref{fig:aia}: relative FIP maps (top) and their corresponding histograms (bottom), with matching colour scales. The DEM inversion code was not able to find a satisfactory solution in the pixels depicted in white.
    The vertical lines in the histograms correspond to the imposed uniform values of the relative FIP bias (for each of the low-FIP elements; see Table~\ref{tab:FIP_bias_ratios}) that should ideally be retrieved.}
    \label{fig:FIP_synthetic}
\end{figure}

\begin{figure}
    \includegraphics[width=\linewidth]{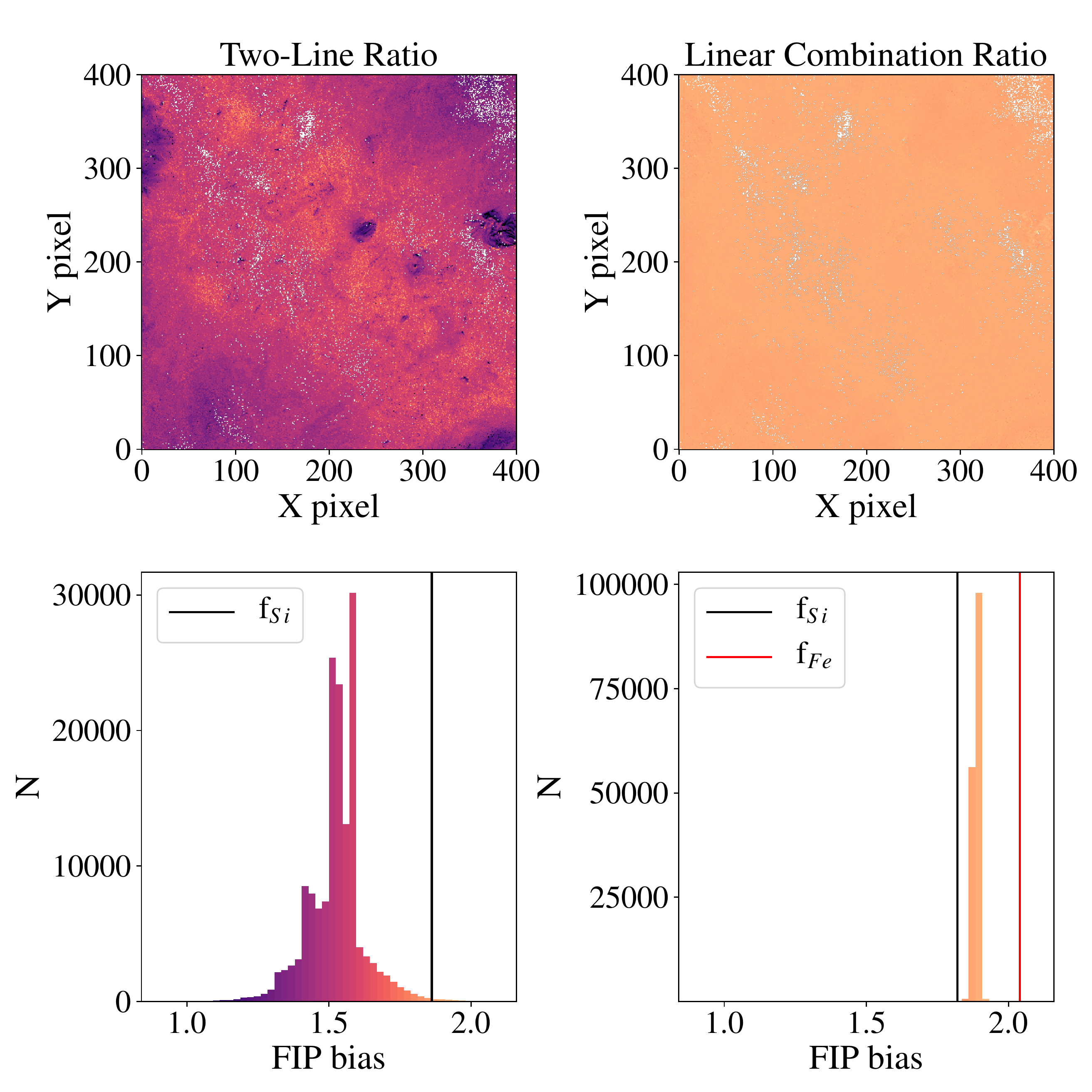}
    \caption{Same as Fig.~\ref{fig:FIP_synthetic} but for the second region of interest (red square) of Fig.~\ref{fig:aia}.}
    \label{fig:FIP_synthetic_ch}
\end{figure}

We present the results for the first region of interest (black square in Fig.~\ref{fig:aia}) in Fig.~\ref{fig:FIP_synthetic}.
The top left panel of this figure clearly shows that we do not retrieve a uniform relative FIP bias using the 2LR method, as confirmed by the standard deviation of the FIP bias (0.15) and the corresponding histogram (bottom left).
Furthermore, the histogram peak at about 1.51 is far from the imposed value for the relative FIP bias between the two elements used, silicon and sulfur (1.82).
This could be because the normalized contribution function of the \ion{S}{x} line goes well over (up to a factor 3.6) that of the \ion{Si}{x} line in the temperature range at which the EM peaks ($\log T = 6.3$ to $6.4$), as we can see in Fig.~\ref{fig:cofnts}.

The LCR method gives a much more uniform map (top right panel), as confirmed by the corresponding histogram (bottom right) that has a standard deviation of 0.03, a factor of five smaller than that obtained with the 2LR method.
Almost all obtained values are between the relative FIP biases for Fe and Si, as we discuss in Sect.~\ref{subsec:nonunif}. This histogram peaks at 1.87.
These results show the accuracy of the linear combination ratio method.

In order to test if these results can be reproduced in regions other than an AR, we perform the same test in the red square of Fig.~\ref{fig:aia}.
We can see in Fig.~\ref{fig:aia} that this region contains very different structures than the first one, as the second region includes part of a CH.
In the results, presented in Fig.~\ref{fig:FIP_synthetic_ch}, we can see that the LCR method performs again better than the 2LR technique.
We obtain a distribution of relative FIP biases peaking at 1.58 (with a standard deviation of 0.1) for the 2LR method, still very far from 1.82, and a distribution peaking at 1.9 (with a standard deviation of 0.015) for the LCR method.
In this case, almost all values are again between the relative FIP biases for Fe and Si.

The LCR results are very close to the imposed FIP biases in both regions even though their EMs are very different (and each region already contains pixels with different EMs).
This shows that the LCR method works properly and does not require prior knowledge of the DEM.

\subsection{Understanding the remaining non-uniformity in maps}
\label{subsec:nonunif}

\begin{figure}
    \includegraphics[width=\linewidth]{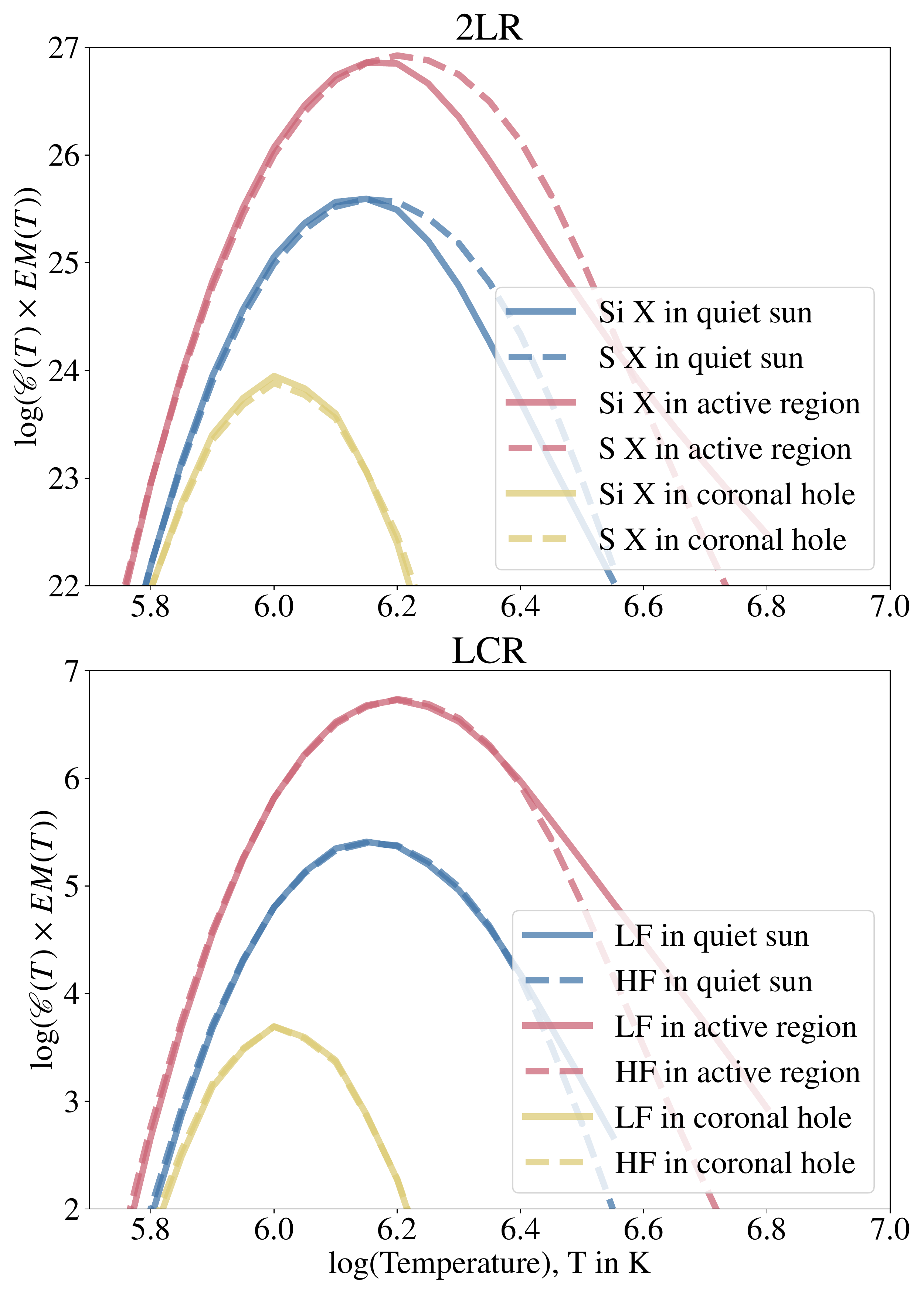}
    \caption{Products between the EM and the contribution functions ($\mathscr{C}_\LF$, solid lines, and $\mathscr{C}_\HF$, dashed lines), as functions of the temperature for both methods, using the coefficients in Table~\ref{tab:spec_lines}.
    Different colours correspond to different EMs from CHIANTI.}
    \label{fig:product_comparison}
\end{figure}

\begin{figure}
    \includegraphics[width=\linewidth]{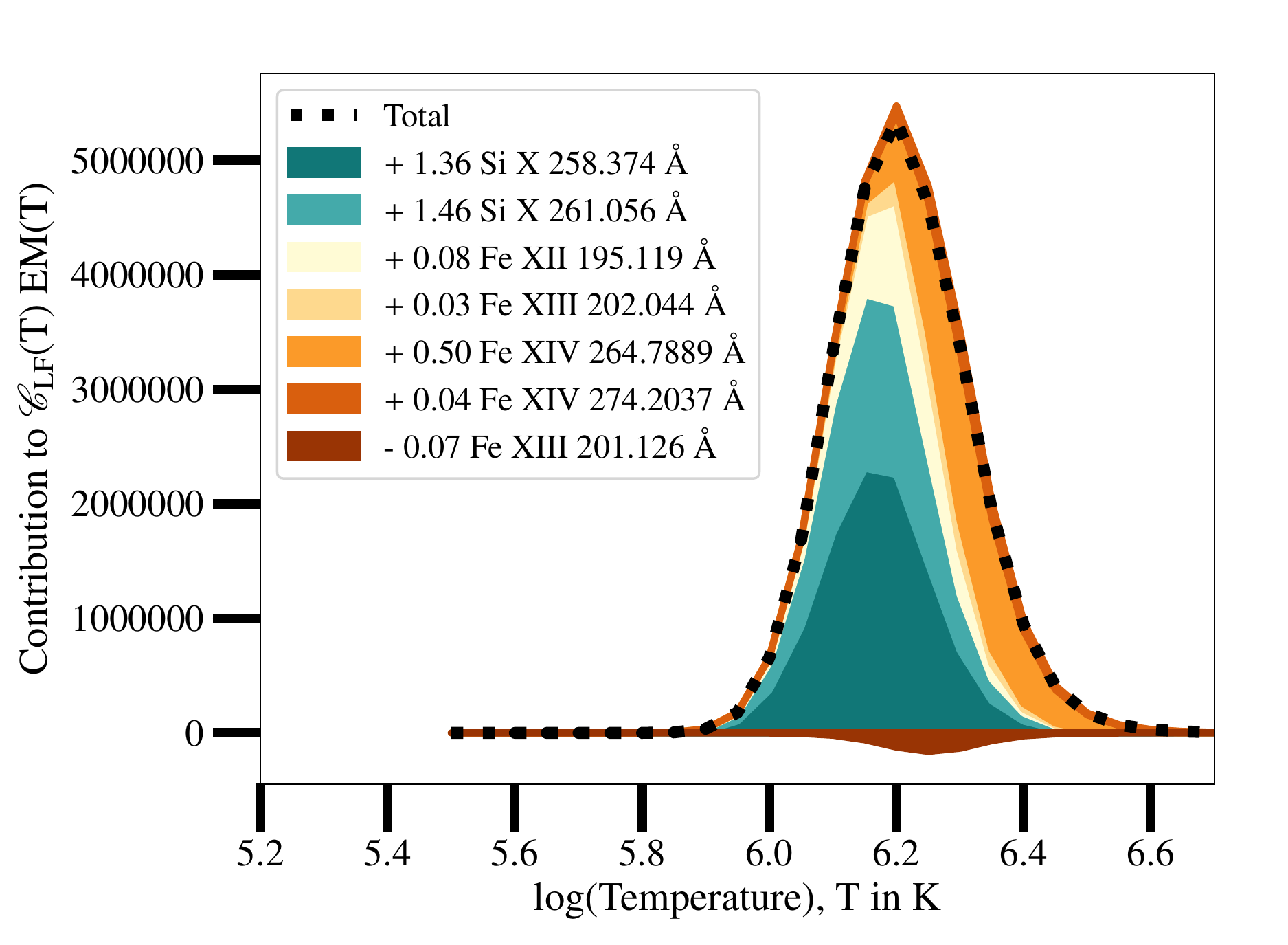}
    \caption{
    Contributions of the different spectroscopic lines to the $\mathscr{C_\LF}(T) \, \EM(T)$ product as a function of temperature for the CHIANTI AR EM.
    The total (dotted line) corresponds to the full red line in the bottom panel of Fig.~\ref{fig:product_comparison}.
    Here, Fe and Si line contributions are shown in ocher-red and blue-green colours, respectively.
    We separate the positive and negative contributions to the total. 
    }
    \label{fig:partial_sum}
\end{figure}

As the assumed FIP bias map (the ground truth for the test) was uniform, the non-uniformity in the result FIP bias map is a measure of the error in the FIP bias given by the tested method (LCR or 2LR) for this test setup (for the spectroscopic lines, reference EMs, and EM map used for the test).
Other sources of error that cannot be assessed with such a test are discussed in Sect.~\ref{sec:discussion}.

Even though the relative FIP bias values obtained in this test with the LCR method have a standard deviation of 0.02 and 0.03 only in both regions of interest, the 1\textsuperscript{st} and 99\textsuperscript{th} percentiles are 1.78 and 1.88 respectively.
Although much better than for the 2LR method, the non-uniformity of the FIP bias is still significant, and we ought to understand possible sources of the remaining non-uniformity in our test FIP bias maps.

\paragraph{Optimisation residuals.}
Even after the optimisation, the residuals of the cost function Eq.~\eqref{eq:costrad} are not zero (Table~\ref{tab:vector_obj_func}).
From Eq.~\eqref{eq:costvec}, one understands that these residuals come from the fact that the products $\mathscr{C}_\LF(T) \, \EM(T)$ and $\mathscr{C}_\HF(T) \, \EM(T)$ for the different EMs used for the optimisation are not close enough.
We trace these products for the three reference EMs we used and for both methods in Fig.~\ref{fig:product_comparison}.

Above $\log T = 6.2$, both \LF\ and \HF\ curves have the same shape for the LCR linear combinations, whereas this is not the case for the 2LR method.
The yellow curves in the top panel of Fig.~\ref{fig:product_comparison} show that, with the AR EM, the 2LR method would overestimate the \ion{S}{X} contribution in the $\log T \in [6.2, 6.6]$ range by a factor up to 3.6. Above $\log T = 6.5$, the \ion{S}{X} contribution would be underestimated for both methods, but we are far from the peak which means that the contribution of the radiance at these high temperatures might not contribute much to the overall observed radiance.

After integration over $T$, the resulting $\langle \mathscr{C}, \EM_\textrm{AR}\rangle$ is higher for S than for Si, corresponding to the fact that $\psi_\textrm{AR}$ is 24\% lower than 1 (see Table~\ref{tab:vector_obj_func}).
This is consistent with the strong underestimation of the FIP bias that is obtained following this test in active regions with the 2LR method, as seen in Figs.~\ref{fig:FIP_synthetic} and~\ref{fig:FIP_synthetic_ch}.

In contrast, for the LCR method, the mean distance between the \LF\ and \HF\ curves in Fig.~\ref{fig:product_comparison} is much smaller than for the 2LR method.
This is measured, after integration over $T$, by the values of $\psi_j$ in Table~\ref{tab:vector_obj_func}: these values are very close to 1.

Overall, this means that the LCR method performs better than the 2LR method in the range of temperatures including the peak of the mean coronal DEM.

\paragraph{Cost function residuals for real DEMs.}
The analysis of the cost function residuals in the previous paragraph is for the set of reference EMs that were chosen for the optimisation.
However, the DEMs in the map are different.
With real observations, we cannot measure the impact of this choice unless we perform a thorough DEM analysis, but this is not an issue in the case of the synthetic observations we produced for our tests in this section.

By applying Eq.~\eqref{eq:costvec} to the EM we used to produce synthetic radiances in every pixel, we can then retrieve the uncertainty linked to the arbitrary choice of reference EMs.
In other words, we can determine through this calculation how far $\psi_{\EM}$ is from one for the EM in every pixel.
As we can see in Eq.~\eqref{eq:biaslcopt}, this factor determines if we over- or underestimate the relative FIP bias.
In both test regions, $\psi_{\EM}$ is $1\pm 0.01$, meaning that in our case the impact on the FIP biases of the fact that the EMs in the map are not those chosen for the optimisation of the LCR coefficients is 1\%.
Therefore, the optimal linear combinations seem to be very well adapted to these EMs even though they were not optimised for them specifically.

\paragraph{Use of different low-FIP elements.}
As most values obtained in the test with the LCR method are between the relative abundance biases of Si (1.82) and Fe (2.05), one reason for the remaining non-uniformity in the maps could be the use of lines of \LF\ elements with different abundance biases, while we assumed from Eq.~\eqref{eq:lincombradratio} that they were the same.
To assess this potential reason, we determine how much the spectral lines of each element are contributing to the total linear combination of \LF\ elements in order to fit as best as possible to our \HF\ line.

In our case, we show in Fig.~\ref{fig:partial_sum} the respective contributions of the Fe and Si lines to the $\mathscr{C}(T) \, \EM(T)$ product for the \LF\ linear combination of the LCR method and for the AR EM (for which the differences in the $\log T \in [6.2, 6.5]$ interval were most noticeable for the 2LR method, as discussed above).
The relative contributions of the Fe and Si lines depend on temperature.
This is true in this case, with the AR EM, but these proportions will vary for different DEMs.
As a result, for any given DEM, the FIP bias given by the LCR method will be closer to that of one element or the other, which can explain a part of the dispersion seen in the histograms of Figs.~\ref{fig:FIP_synthetic} and~\ref{fig:FIP_synthetic_ch}.

\section{Determining FIP bias from observations} \label{sec:baker_lines}

We applied the LCR method to spectroscopic observations of a sigmoidal anemone-like AR inside an equatorial CH that has previously been studied (including plasma composition) in \cite{Baker13}.
A full description of the evolution of this AR from the 11\textsuperscript{} to the 23\textsuperscript{} October, 2007, including measurement of multi-temperature plasma flows, is presented in \cite{Baker2012}.
We focus on a single raster observation lasting 2.25 hours that was carried out with the EIS spectrometer \citep{Culhane07} aboard Hinode \citep{Kosugi07} on October 17, 2007, at 2:47\,UT.

\citet{Baker13} used the method described in Appendix~\ref{app:dem_inversion} in order to retrieve FIP bias maps.
These latter authors used ten Fe lines in order to infer the EM from line radiances.
They scaled this EM to accurately reproduce the radiance of the same \ion{Si}{x} line that we used previously for the 2LR method (see Table~\ref{tab:spec_lines}). They then simulated the radiance of the same \ion{S}{x} line that we used in the previous section and compared it to the observed radiance. The ratio gives a FIP bias map, reproduced in the left panel of Fig.~\ref{fig:baker_vs_lc}.

In our analysis, we start by applying standard SolarSoftware EIS data-reduction procedures to the data, including correcting for dark current hot, warm, and dusty pixels, cosmic rays, slit tilt, CCD detector offset, and orbital variations. The obtained calibrated spectra were then fitted by single (or double, when necessary) Gaussian functions, and we computed integrated radiances for all lines.

\begin{figure}
    \includegraphics[width=\linewidth]{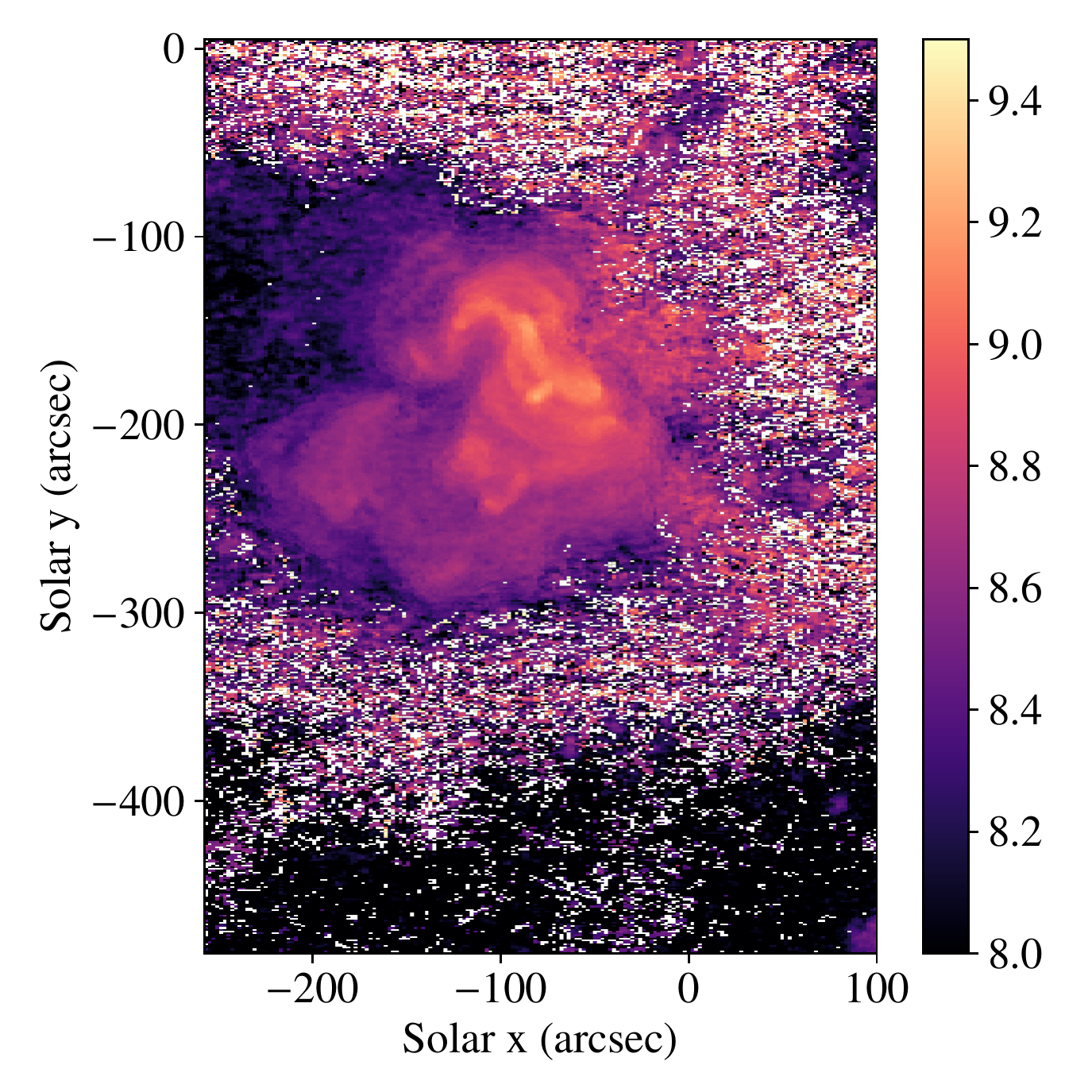}
    \caption{Map of the logarithm of the density in cm$^{-3}$ obtained for the observation on October 17,\textsuperscript{} 2007, from EIS spectra. We used the \ion{Fe}{xiii} $\lambda$ 202.02 and 203.83 line pair to calculate this density map. In the white pixels, fitting of the EIS lines failed. The density value of each pixel allows us to compute the optimal linear combination of lines to be used to determine the FIP bias in those pixels.}
    \label{fig:density}
\end{figure}

\begin{figure}
    \includegraphics[width=\linewidth]{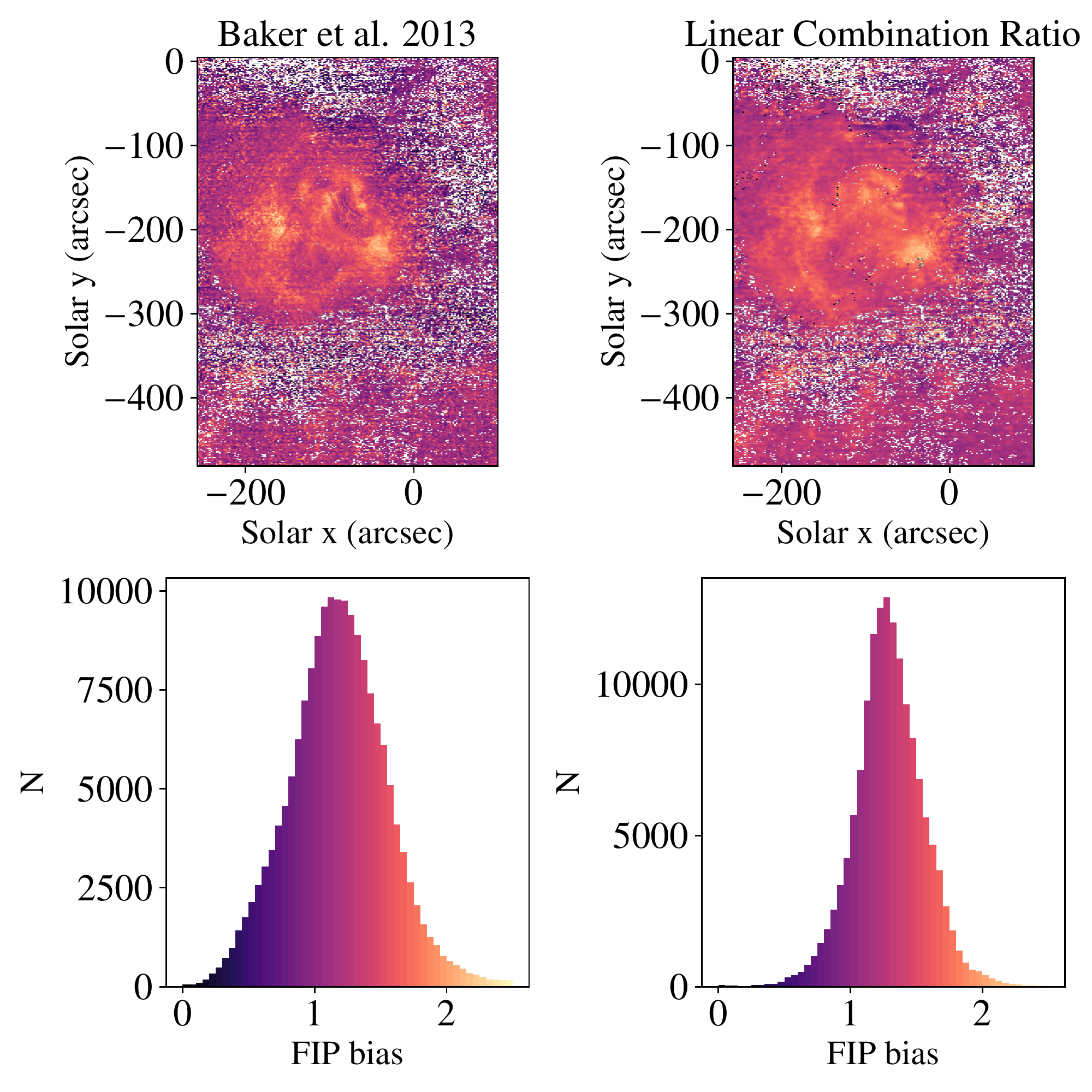}
    \caption{First ionization potential bias maps obtained with different methods and the corresponding histograms. The left panel shows the FIP bias map obtained following DEM inversion (adapted from \citealt{Baker13}). The right panel shows the FIP bias map obtained using the LCR method. In the white pixels, the fitting of the EIS lines we performed failed. In both FIP bias maps we only plot the pixels where our fitting was successful.}
    \label{fig:baker_vs_lc}
\end{figure}

We then selected the lines to be used for the LCR method, using the criteria from Sect.~\ref{subsubsec:line_selection}.
This gives the five Fe lines, the two Si lines, and the S line listed in Table~\ref{tab:spec_lines}.
We calculated the density of this AR using the \ion{Fe}{xiii} $\lambda$ 202.02 and 203.83 line pair diagnostic. This density map is plotted in Fig.~\ref{fig:density}. We then determined the optimal coefficients to use in each pixel using this map.
In the case of this EIS observation, all the selected lines have strong radiances and are fairly isolated in the spectrum.
However, among the EIS windows of this observation, only one line of an element considered as \HF\ fits all our selection criteria.
As a result, the set of \HF\ lines is reduced to a single line (as in Sec.~\ref{sec:tests}).

\begin{figure}
    \includegraphics[width=\linewidth]{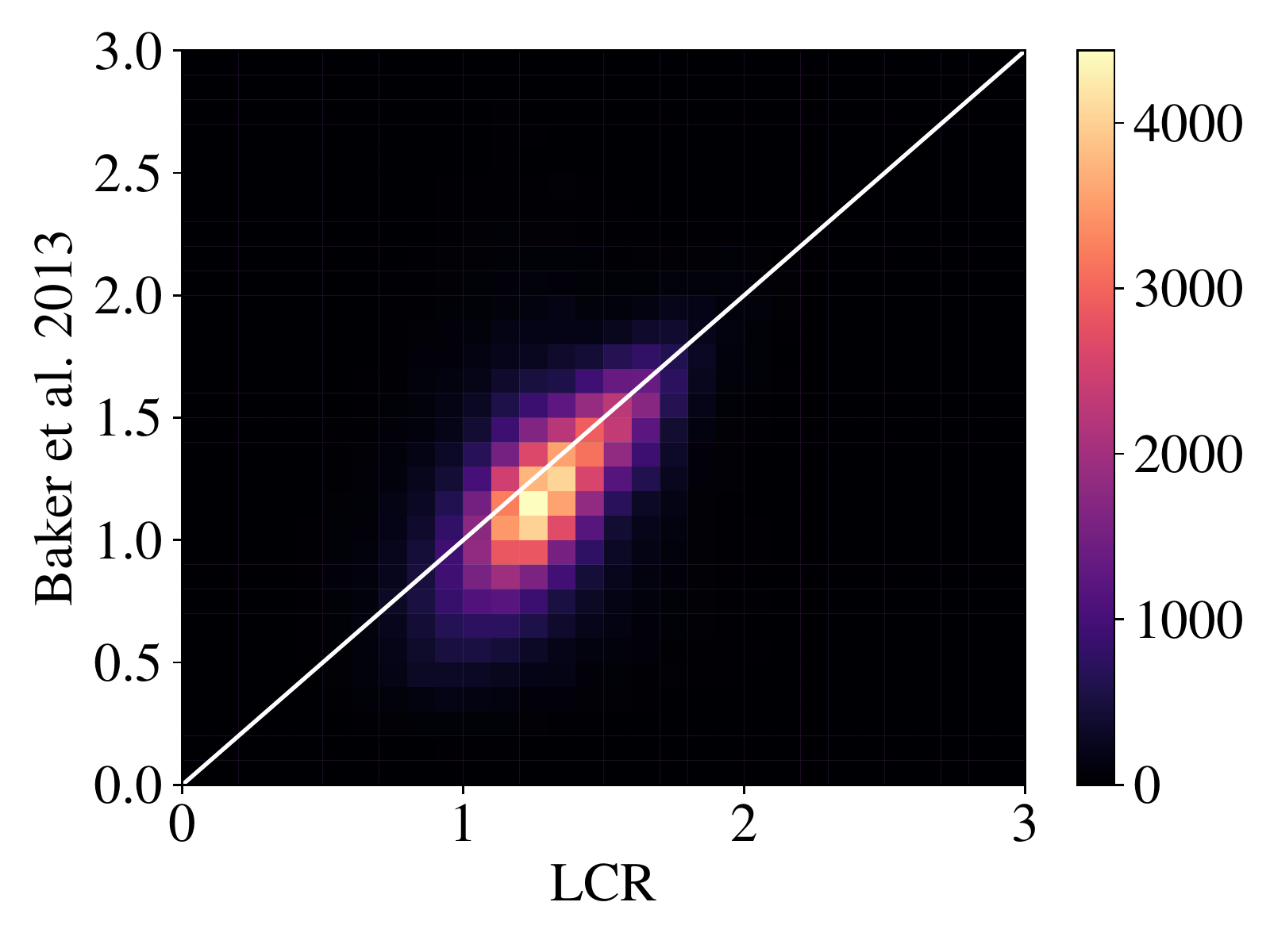}
    \caption{Two-dimensional histogram of the FIP bias values of \citet{Baker13} and those obtained using the LCR method. The white line is the first bisector, where both values are equal.}
    \label{fig:hist2d}
\end{figure}

The results of the LCR method on this observation are presented in the right panel of Fig.~\ref{fig:baker_vs_lc}.
The FIP bias maps (from \citealt{Baker13} and from the LCR method) display similar FIP bias structures.
The distributions of FIP bias values (bottom panels of Fig.~\ref{fig:baker_vs_lc}) peak at 1.11 and 1.29.
The correlation between both sets of values (Fig.~\ref{fig:hist2d}) also shows that the LCR values are higher overall than the \citet{Baker13} values.
However, we do not expect a perfect correlation as the real FIP bias values in this region are not known.
We find that the LCR FIP bias map provides useful information on the FIP biases in the coronal structures in the field of view, which is remarkable given that it was produced without any DEM inversion

\section{Discussion} \label{sec:discussion}

Some sources of errors that could be identified from the non-uniformity in the test result in Sect.~\ref{sec:tests} have already been discussed in Sect.~\ref{subsec:nonunif}: the cost function residuals for the reference DEMs and for the real DEMs in the map, and the assumption that all \LF\ (or \HF) elements used have the same abundance bias.
The cost function residuals for the real DEMs could be reduced by using a more comprehensive set of reference DEMs, or a set that would be more adapted to the observation; but the latter would require better knowledge of the DEMs in the observation, and we wanted to avoid inverting DEMs.
The tests we have done show however that the residuals are in practice small for real DEMs given the set of three reference DEMs that we have chosen for the optimisation, and so this set is sufficient.

In regards to the mixing, in the same group (\LF\ or \HF) of spectroscopic lines from different elements with different abundance biases this is a matter of compromise.
As one can see in Fig.~\ref{fig:partial_sum}, for the set of lines that we have chosen (the same as the ones available in the observation analysed by \citealt{Baker13} and re-analysed in Sect.~\ref{sec:baker_lines}), using only Si lines would not have allowed us to fit the \LF\ and \HF\ $\mathscr{C}_\LF(T) \, \EM(T)$ products, especially in the most relevant temperature interval (close to the DEM peak).
The Fe lines provide a better fit, and subsequently a smaller value for the optimised cost function.
This gives in the end a more accurate FIP bias determination, although the assumption that all abundance biases are the same for all \LF\ or \HF\ elements has not been verified; this must be checked on a case-by-case basis, depending on the elements giving the available spectral lines, along with their behavior with respect to the FIP effect.

As in any UV spectroscopic analysis, other uncertainties come from radiometry (inaccurately measuring the line radiances, e.g. because of calibration or line blends), atomic physics (imprecise atomic data for computing the contribution functions; not taking into account effects such as those from non-Maxwellian distributions or from non-equilibrium of ionization, when required) and radiative transfer (opacity and scattering).

\section{Conclusion}

Here, we present the LCR method, developed with the aim to provide optimal determination of the relative FIP biases in the corona from spectroscopic observations without the need to previously determine the DEM.
This technique relies on linear combinations of spectral lines optimised for FIP bias determination. We developed a Python module to implement the method  that can be found at \url{https://git.ias.u-psud.fr/nzambran/fiplcr}.

Using two linear combinations of spectral lines, one with low FIP elements and another with high FIP elements, we tested the accuracy of the method performed on synthetic observations: these tests show that the method does indeed perform well, without prior DEM inversions.
We then applied it to Hinode/EIS observations of an active region. We obtained FIP bias structures similar to those found in the same region by \cite{Baker13} following a DEM inversion.

Once the optimised linear combination coefficients have been determined for a given set of lines, if radiance maps can be obtained in these lines, the LCR method directly gives the corresponding FIP bias maps, in a similar way to the 2LR method, but with better accuracy.
This makes the method simple to apply on observations containing a pre-defined set of lines, with a potential for automation.

Hopefully, producing such FIP bias maps semi-automatically will allow non-specialists of EUV spectroscopy to obtain composition information from remote-sensing observations and compare it directly with in-situ data of the solar wind.
This method could also allow better exploitation of observations not specifically designed for composition studies, and an optimal design of future observations.
We plan to apply the method to the future Solar Orbiter/SPICE spectra to prepare the observations and analysis of the SPICE data.

\begin{acknowledgements}
The authors thank Mark Cheung for the SDO/AIA DEM inversion code, Deborah Baker for providing the FIP bias map data, and Giulio Del Zanna, Susanna Parenti, and Karine Bocchialini for their comments.
    The idea for this work came following a presentation by Hardi Peter at the Solar Orbiter joint SPICE-EPD-SWA meeting in Orsay in November 2015.
    AIA is an instrument on board SDO, a mission for NASA's Living With a Star program.
    NZP thanks ISSI for support and participants to the ISSI team n. 418 "Linking the Sun to the Heliosphere using Composition Data and Modelling" led by Susanna Parenti.
    Hinode is a Japanese mission developed and launched by ISAS/JAXA, with NAOJ as domestic partner and NASA and STFC (UK) as international partners. It is operated by these agencies in cooperation with ESA and NSC (Norway).
    CHIANTI is a collaborative project involving George Mason University, the University of Michigan (USA) and the University of Cambridge (UK).
    This work used data provided by the MEDOC data and operations centre (CNES / CNRS / Univ. Paris-Sud), \url{http://medoc.ias.u-psud.fr/}.
    Python modules used include: \texttt{numpy}, \texttt{matplotlib}, \texttt{astropy}, \texttt{scipy}, \texttt{sunpy}, \texttt{chiantipy} and \texttt{colorblind} (available at \url{https://github.com/volodia99/colorblind}).
\end{acknowledgements}

\bibliographystyle{aa}
\bibliography{sample}

\appendix

\section{Using DEM inversion to derive FIP bias}
\label{app:dem_inversion}

We describe here the general idea behind the method used by \cite{Baker13,Guennou2015}. Their FIP bias determination relies in the following steps:
\begin{itemize}
    \item Retrieve radiances from observations of a number of spectral lines from low FIP and high-FIP elements.
    \item Determine the density (for every pixel in the observation) and use it to compute the contribution functions of the spectral lines used for the analysis.
    \item Infer the DEM from the radiances of the spectral lines of a low FIP element only, assuming {photospheric} abundances. This `inferred' DEM is obtained by inversion of the observed radiances written as
    \begin{equation}
        I_{ij, X_\LF} = A^\phot_{X_\LF} \, \langle C_{ij, X_\LF}, \DEM^\textrm{inferred}\rangle.
        \label{eq:assumed_int}
    \end{equation}
    As in reality this element is subject to the FIP effect, the radiances are in fact
    \begin{equation}
        I_{ij,X_\LF} = f_{X_\LF} \,A^\phot_{X_\LF} \, \langle C_{ij, X_\LF}, \DEM\rangle,
        \label{eq:real_int}
    \end{equation}
    where the DEM is the real DEM.
    The inferred DEM is then overestimated by a factor
    \begin{equation}
        \frac{\DEM^\textrm{inferred}}{\DEM} = f_{X_\LF}.
    \end{equation}
    \item Compute the ratio of the simulated (with $\DEM^\textrm{inferred}$)\footnote{If one wants to compute the FIP bias for another low-FIP element (e.g. Si, when the DEM was computed using Fe lines), the inferred DEM is rescaled so that it reproduces the observed radiance of a line of that element.} and observed radiances of a high-FIP element spectral line:
    \begin{align}
        \frac{I^\textrm{simulated}_{X_\HF}}{I^\textrm{observed}_{X_\HF}} &=
        \frac{A^\phot_{X_\HF} \, \langle C_{X_\HF}, \DEM^\textrm{inferred}\rangle}{f_{X_\HF} \, A^\phot_{X_\HF} \, \langle C_{X_\HF}, \DEM\rangle} \\
        &= \frac{f_{X_\LF} \, A^\phot_{X_\HF} \, \langle C_{X_\HF}, \DEM\rangle}{ f_{X_\HF} \, A^\phot_{X_\HF} \, \langle C_{X_\HF}, \DEM\rangle} \\
        &= \frac{f_{X_\LF}}{f_{X_\HF}}.
    \end{align}
    This ratio is then the relative FIP bias.
\end{itemize}

\section{Density dependence}
\label{app:density_independance}

\begin{figure}
    \includegraphics[width=\linewidth]{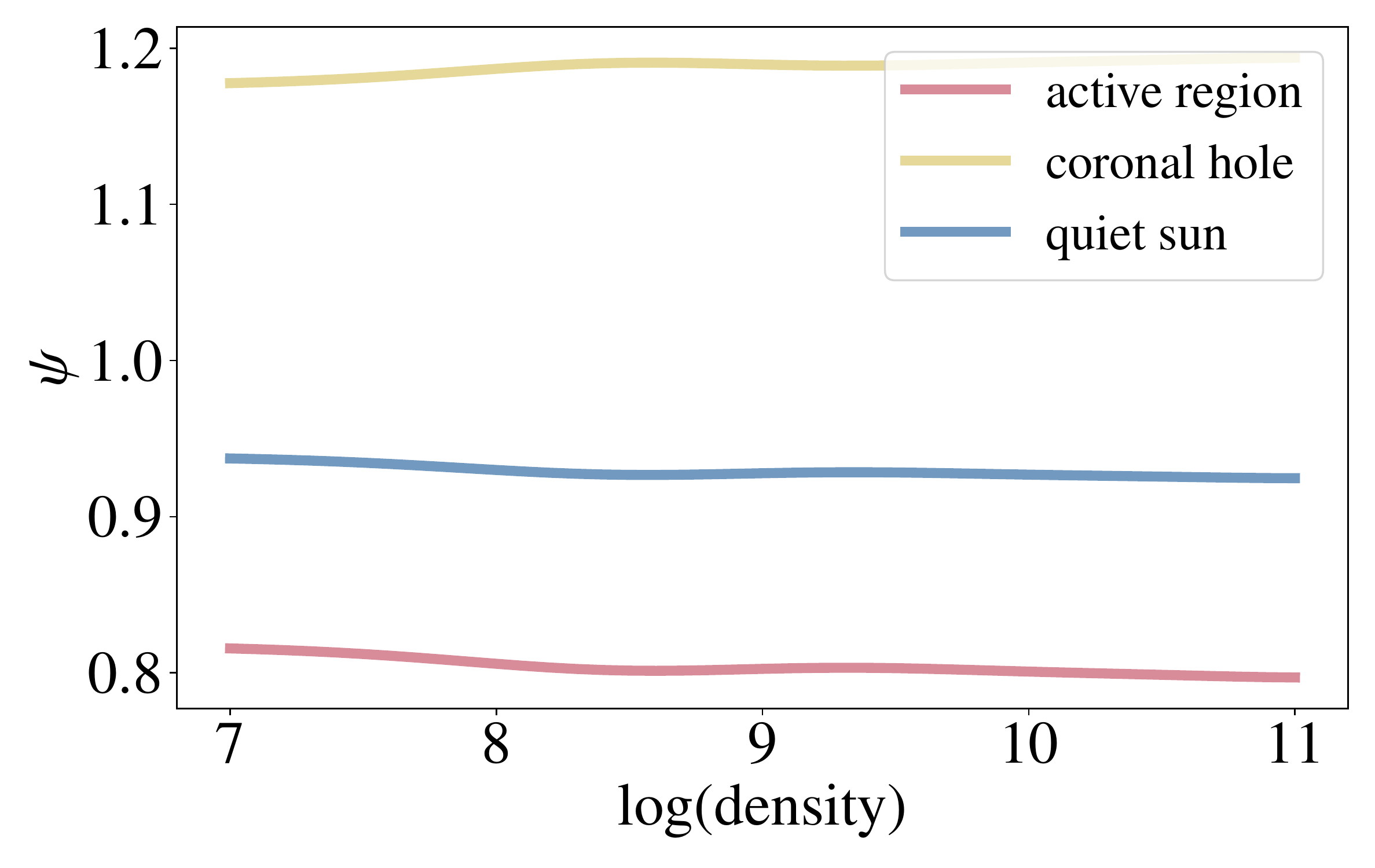}
    \caption{Value of $\psi$ from Eq.~\eqref{eq:costvec} as a function of density for each typical EM from CHIANTI for the lines used in the 2LR method. The coefficients used to calculate $\psi$ for every density value are defined by Eq.~\eqref{eq:2LR_coeffs}. The black lines correspond to the values of $\log n = 8.3$ for the abscissa and $\psi = 1$ for the ordinates.}
    \label{fig:psi_2lr}
\end{figure}

\begin{figure}
    \includegraphics[width=\linewidth]{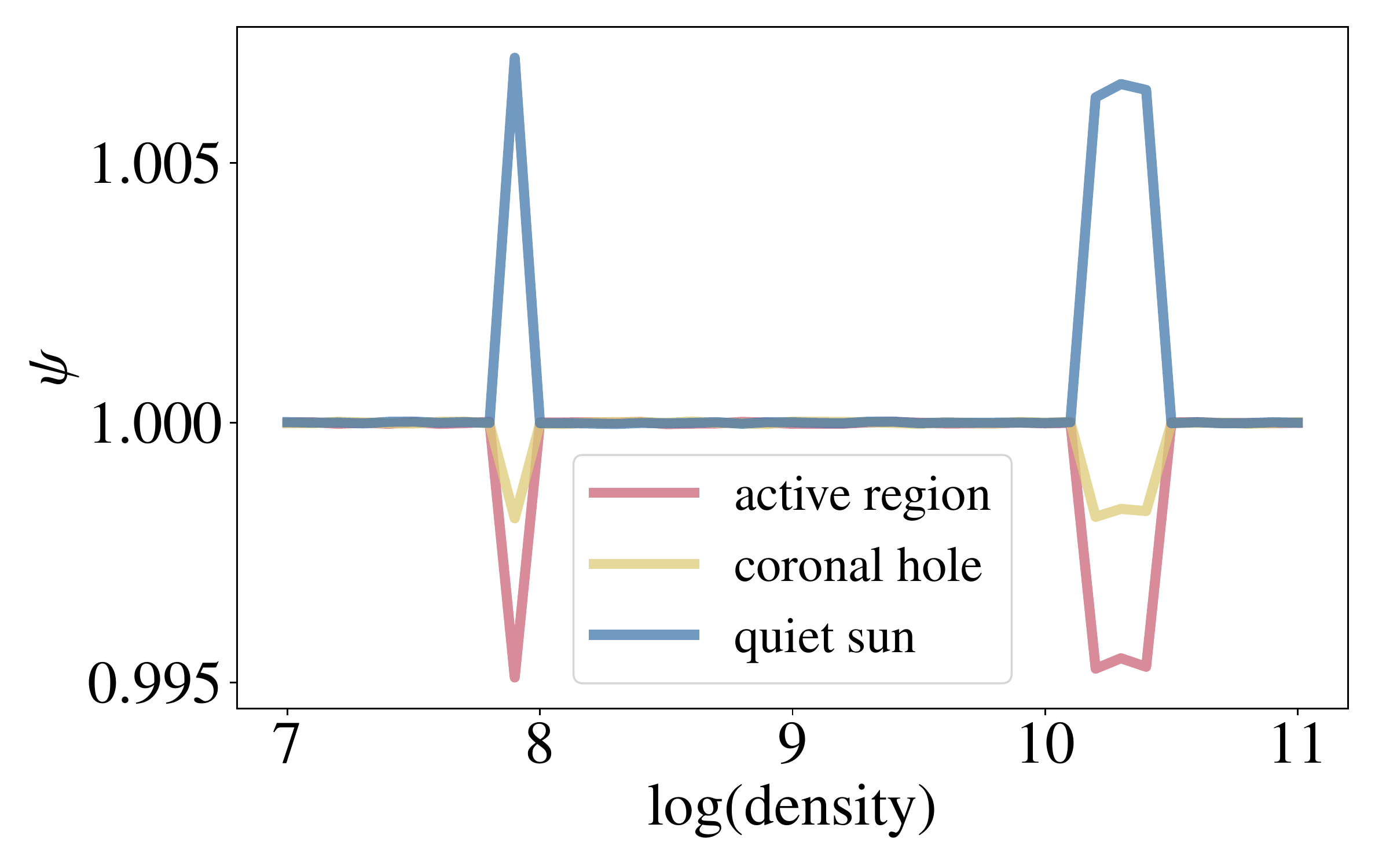}
    \caption{Value of $\psi$ from Eq.~\eqref{eq:costvec} as a function of density for each typical EM from CHIANTI for the linear combinations of lines used in the LCR method. The coefficients used to calculate $\psi$ for every density value are the result of the optimisation. For example, the coefficients used for every line at log(n)=8.3 are the ones listed in Table~\ref{tab:spec_lines}.}
    \label{fig:psi_lcr}
\end{figure}

As mentioned in Sect.~\ref{sec:optimlc}, following the density-dependence of the contribution functions, the coefficients of the linear combinations also depend on density.
We traced the resulting value of $\psi$ as a function of density from Eq.~\eqref{eq:costvec} for the 2LR method in Fig.~\ref{fig:psi_2lr} and for the LCR method in Fig.~\ref{fig:psi_lcr}.
We perform the calculations using the three typical EMs from CHIANTI mentioned above and plotted in Figs.~\ref{fig:em_histo_ar} and~\ref{fig:em_histo_ch}.
The variable $\psi$ represents the ratio of the radiances of the linear combinations of spectral lines if the FIP biases would be 1.
The goal of the optimisation in the LCR method is to have $\psi$ be as close to 1 as possible, so that the relative FIP bias is given by the ratio of the linear combinations of spectral lines as defined in Eqs.~\eqref{eq:pseudo_intensity} to~\eqref{eq:lincombradratio}.

As we can see in Fig.~\ref{fig:psi_2lr} (and consistent with the values of Table~\ref{tab:vector_obj_func} at $\log n = 8.3$), the 2LR method gives values of $\psi$ that can be up to 20\% above or below the target value of 1, leading to erroneous FIP bias determination. The $\psi$ value for each EM depends somewhat on density as well.
In contrast, the LCR method (Fig.~\ref{fig:psi_lcr}) yields $\psi$ values that are less than 0.7\% away from 1 at all densities, leading to much more accurate FIP bias determination than with the 2LR method.

\begin{figure}
    \includegraphics[width=\linewidth]{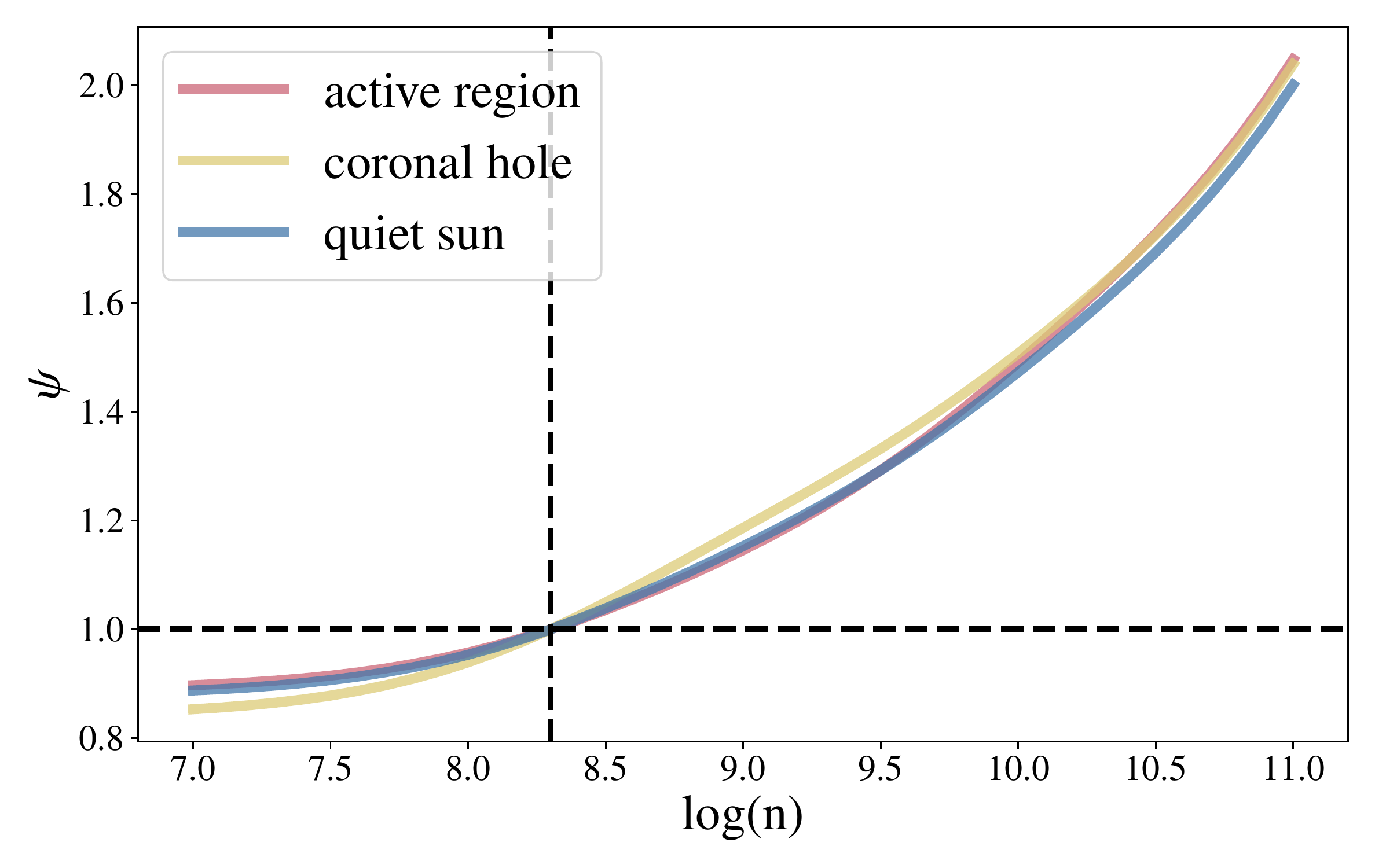}
    \caption{Value of $\psi$ from Eq.~\eqref{eq:costvec} as a function of density for each typical EM from CHIANTI for the linear combinations of lines used in the LCR method. This time we use the coefficients listed in Table~\ref{tab:spec_lines} computed at a fixed density of $\log n = 8.3$ and the contribution functions are evaluated at the different density values. The dashed lines correspond to the density value assumed for the optimisation, and to the target value of $\psi$.}
    \label{fig:psi_const_dens}
\end{figure}

We then trace $\psi$ in Fig.~\ref{fig:psi_const_dens} using only the coefficients computed for the LCR method at a fixed density of $\log n = 8.3$ and the contribution functions evaluated at different density values. 
We do so to determine the error one would commit by assuming a constant density of $\log n = 8.3$ when determining the optimised coefficients (listed in Table~\ref{tab:spec_lines}) instead of using the density-dependent approach.
In this case, $\psi$ remains within 20\% of the target value of 1 from below $\log n = 7$ to $\log n = 9$, meaning that the LCR method can perform as well as or better than the 2LR method in a significant range of densities, even when not taking the density dependence of the optimal coefficients into account.
However, $\psi$ deviates strongly from 1 for higher densities, up to a factor two if $\log n = 11$ instead of $8.3$.

\end{document}